\documentclass[12pt]{article}
\usepackage{natbib}
\usepackage{url} 
\usepackage{amsmath, amssymb}
\usepackage{graphicx}

\newcommand{\blind}{0}

\usepackage{hyperref}
\usepackage[dvipsnames, table]{xcolor}
\hypersetup{
    colorlinks,
    linkcolor={red!50!black},
    citecolor=NavyBlue,
    urlcolor={blue!80!black}
}

\usepackage{tikz}
\usetikzlibrary{arrows.meta, positioning, quotes, matrix}
\usepackage[mode=buildnew]{standalone}

\renewcommand{\Pr}{\text{Pr}}
\newcommand{\E}{\mathbb{E}}
\newcommand{\V}{\mathbb{V}}

\newcommand{\II}[1]{\mathbb{I} \left\{#1\right\}}   
\newcommand{\IIrep}{\mathbb{I}^{(i)}}                      
\newcommand{\ppp}{\mbox{ppp}}                           
\newcommand{\ppph}{\widehat{\ppp}}                    
\newcommand{\trep}{\theta^{(i)}}     
\newcommand{\yrep}{y^{*(i)}}            
\newcommand{\yc}{\tilde{y}}

\newcommand{\Yc}{\widetilde{Y}}
\newcommand{\kc}{\tilde{k}}
\newcommand{\Kc}{\widetilde{K}}
\newcommand{\IIci}{\tilde{\mathbb{I}}^{(i)}}

\newcommand{\tauc}{\tilde{\tau}} 
\newcommand{\ycrep}{\tilde{y}^{(j)}}            
\newcommand{\kcrep}{\tilde{k}^{(j)}}            
\newcommand{\mcrep}{\tilde{m}^{(j)}}            
\newcommand{\mc}{\tilde{m}}            

\newcommand{\ESS}{\mbox{ESS}}               
\newcommand{\ESSc}{\widetilde{\mbox{ESS}}}  

\newcommand{\cppp}{\mbox{cppp}}                  
\newcommand{\cppph}{\widehat{\cppp}}           

\usepackage{cellspace}
\usepackage{multirow}

\usepackage{makecell}

\newcommand{\inlineRevised}[1]{\textcolor{black}{#1}}

\begin{document}

\def\spacingset#1{\renewcommand{\baselinestretch}%
{#1}\small\normalsize} \spacingset{1}

\if0\blind
{
  \title{\bf Computational methods for fast Bayesian model assessment via calibrated posterior p-values}
  \author{Sally Paganin\thanks{\textit{This work was partially funded by US National Science Foundation Grant No. 1550488. The authors would like
to thank Chris Paciorek and Massimiliano Russo for their suggestions on earlier versions of the manuscript. }}\hspace{.2cm}\\
    Department of Statistics, The Ohio State University\\
    and \\
    Perry de Valpine\\
    Department of Environmental Science, Policy, and Management,\\ University of California, Berkeley}
    \date{}
  \maketitle
} \fi

\if1\blind
{
  \bigskip
  \bigskip
  \bigskip
  \begin{center}
    {\LARGE\bf Title}
\end{center}
  \medskip
} \fi

\bigskip
\begin{abstract}
\inlineRevised{Posterior predictive p-values (ppps) have become popular tools for Bayesian model assessment, being general-purpose and easy to use. However, interpretation can be difficult because their distribution is not uniform under the hypothesis that the model did generate the data. 
Calibrated ppps (cppps) can be obtained via a bootstrap-like procedure, yet remain unavailable in practice due to high computational cost. This paper introduces methods to enable efficient approximation of cppps and their uncertainty for fast model assessment. We first investigate the computational trade-off between the number of calibration replicates and the number of MCMC samples per replicate. Provided that the MCMC chain from the real data has converged, using short MCMC chains per calibration replicate can save significant computation time compared to naive implementations, without significant loss in accuracy. We propose different variance estimators for the cppp approximation, which can be used to confirm quickly the lack of evidence against model misspecification. As variance estimation uses effective sample sizes of many short MCMC chains, we show these can be approximated well from the real-data MCMC chain. The procedure for cppp is implemented in NIMBLE, a flexible framework for hierarchical modeling that supports many models and discrepancy measures.}
\end{abstract}

\noindent%
\vfill

\newpage
\spacingset{1.5} 
\section{Introduction}
\label{sec:intro}
An important step in statistical modeling is to ask whether a model fits the data reasonably.  In applied Bayesian statistics using Markov chain Monte Carlo (MCMC) algorithms, a common method is the use of posterior predictive checks, which compare some function of the data to a reference distribution, such as the posterior predictive distribution \citep{rubin1984,meng1994posterior,gelman1996posterior}. 
The result of this comparison is typically summarized by a ``p-value'', which serves as an assessment of whether, for some chosen measure of disagreement between model and data, the data are so unusual as to doubt the veracity of the model.
These posterior predictive p-values (hereafter, ppps) are considered a measure of ``surprise'', with low values indicating that the data are incompatible with the model. 

Unfortunately, it is well known that posterior predictive checks are liberal in the sense of concluding that models are better than they really are. Additionally, posterior predictive p-values lack the frequentist property of following a uniform distribution if the model did generate the data, clustering instead around a value of 0.5 \citep{sinharay2003posterior, gelman2013two}. 
\inlineRevised{As a consequence, high values of the ppp can obtained even in cases of severe mismatch between the data and a model \citep{bayarri2000p}.} 
This makes ppps difficult to interpret and leaves this common practice of Bayesian model assessment on shaky ground. The ppp is a measure of ``surprise'' in units that sound like probability, in the same sense as a frequentist p-value, but are not. \inlineRevised{We refer to \cite{robins2000asymptotic} for a comprehensive discussion on the asymptotic properties of ppps.}

For this reason, many authors have discussed the need for calibration of ppps to set an interpretable scale \citep{robins2000asymptotic,hjort2006post,steinbak2009}. In particular, \cite{hjort2006post} propose calibration of ppps via a bootstrap-like procedure: simulate many data sets from the reference distribution and go through the steps of MCMC and calculation of ppp to obtain a simulated null distribution of ppps. From this null distribution, the frequentist p-value of the Bayesian ppp can be determined, providing a more standardized (or objective) measure of model plausibility. Since the goal is frequentist calibration of a Bayesian method, we speak informally of a null distribution and clarify this notion in Section~\ref{sec:sec2}.

Calibrated ppps (cppps) are appealing because they can provide an automated method for a large class of problems with few additional assumptions. However, the major difficulty with cppps is the potentially prohibitive computational burden of obtaining them. It is not uncommon for analysts to run MCMC for hours or days, so repeating this step a large number of times can be daunting in practice. \inlineRevised{To fill this gap between theory and practical application, this paper introduces computational methods that enable useful approximation of cppps much faster than would be achieved by a naive implementation. Our approximate cppps are based on two key ideas: (i) the use of short MCMC chains for each bootstrap sample (calibration replicate) and (ii) the use of a ``transfer estimate'' to quantify uncertainty in the approximation.}

\inlineRevised{Literature on efficient computation of the cppp is limited. Some approaches \citep{johnson2007bayesian,yuan2012goodness} circumvent calibration by using measures of discrepancy based on pivotal quantities, i.e. statistics of data, parameters, or both, having a known probability distribution under the null hypothesis that is independent of unknown parameters. However, these methods are restricted to particular choices of models and discrepancies.
A more general approach has been proposed in \cite{nott2018approximation} to approximate (calibrated) predictive p-values using regression adjustment approximate Bayesian computation (ABC) in cases where high accuracy of computation is not required. However, the authors focus on the case where the reference distribution is the prior predictive. More recent work focuses on alternative definitions of predictive checks (\citep{li2022split,moran2023holdout}) that build on the idea of splitting the data into training and held-out test sets, avoiding the double use of data. These methods have been proposed as complementary checks to the cppps. } 

Posterior predictive checks remain a model-generic method, which makes no assumptions of analytic tractability and can work for any discrepancy measure of interest. 
\inlineRevised{Calibration of the cppp allows similar interpretation as for frequentist p-values, such as comparison to a chosen threshold (Type I error rate) for statistical significance (e.g., 0.05 or 0.10).}
To achieve fast approximation, the core computational trade-off is between the number of calibration replicates \inlineRevised{and the number of MCMC samples for each replicate used to approximate its ppp.} Guided by theoretical examination of bias and variance in the cppp estimator, we show that only a small MCMC sample for each calibration replicate is necessary.  MCMC samples with effective sample size (ESS)---of the indicator used in ppp calculation---between $50$ and $200$ can be enough to make bias negligible.  Beyond that, it is optimal to spend computational effort on more calibration replicates to reduce the variance of the cppp estimate. \inlineRevised{We also suggest a default starting number of calibration replicates of 50-100. If the resulting confidence interval for the cppp is too large to reach a clear conclusion, more replicates can be added. In our empirical experiments, once the bias is negligible, we found that increasing replicates beyond $1,000$ would rarely be of interest}.

Critically, in the case of a good model fit, even a fairly rough (fast) estimate of cppp may be adequate to conclude the model is acceptable. For example, a report that the cppp is 0.4 with 95\% Monte Carlo confidence interval from 0.3-0.5 would be rough but adequate to conclude the model is acceptable. Therefore, it is important to estimate the Monte Carlo standard error of the cppp estimate itself. However, this requires an estimate of the ESS of the MCMC sample for each calibration replicate. And that creates a conundrum because the ESS estimates from short MCMC chains will themselves be inaccurate, hindering precise estimation of the cppp uncertainty.  To address this, we propose a ``transfer'' approach that uses the properties of the MCMC chain from the real data as a basis for approximating ESS \inlineRevised{of the MCMC samples of the many calibration replicates.}  The chain for the real data will be much longer and hence more informative about mixing properties. As this step is somewhat heuristic, one can build in a conservative buffer. This makes it possible to estimate the cppp and its Monte Carlo uncertainty from short MCMC runs for many calibration replicates. 

Together, these two advances -- using short MCMC chains for each calibration replicate and using few calibration replicates when a rough estimate of the cppp is sufficient -- can save one or more orders of magnitude of computational effort compared to a naive implementation to compute the cppp.  By a ``naive implementation'', we mean one where the analyst runs an MCMC for each calibration replicate that is as long as the MCMC for the real data and uses an unnecessarily large number of calibration replicates.

The rest of the paper is organized as follows.  In Section \ref{sec:sec2}, we set up the ppp and cppp problems.  In Section \ref{sec:sec3}, we give the ppp and cppp estimators and develop expressions for the bias and variance of the latter.  In Section \ref{sec:sec4}, we study the bias-variance trade-off of allocating computational effort between the number of calibration replicates and the number of MCMC samples per calibration replicate. The theory from Section \ref{sec:sec3} suggests that the entire cppp estimation process can be understood as similar to a beta-binomial estimation problem, which we use to give theoretical results on the bias-variance trade-off. Section \ref{sec:sec5} discusses the estimation of the standard error of cppp estimates, including plug-in and bootstrap estimates that use the heuristic proposal to transfer mixing properties of the long MCMC chain to the many short calibration MCMC chains.  Section \ref{sec:sec6} gives two worked examples, one from physics and one from ecology, with scenarios of models that should be accepted and rejected. The procedure for cppp is implemented using the NIMBLE software \citep{nimble-software:2022,devalpine2017nimble} a flexible R-based system for hierarchical models, and code to reproduce examples in the paper is available at 
\url{https://github.com/salleuska/fastCPPP}.

\section{Model-checking, $\ppp$ and $\cppp$}\label{sec:sec2}
Suppose data $y$ are modeled as having been generated from $p(y | \theta)$, with $\theta$ including unknown parameters and latent states. Typically some prior distribution $\pi(\theta)$ is assumed, and $\theta$ is estimated in a Bayesian framework by its posterior $p(\theta| y)$.  Bayesian model checking focuses on assessing whether or not the assumed model, defined as the combination of data distribution $p(y|\theta)$ and prior $\pi(\theta)$, is compatible with data. Informally, we look for evidence against the null hypothesis that the posited model did really generate the data. We refer to \cite{robins2000asymptotic} for a rigorous definition of the testing problem in this context, and analysis of asymptotic properties of Bayesian p-values.
For model-checking, a discrepancy measure $D(y, \theta)$ is chosen, reflecting aspects of the data that the model must describe well. The term \emph{discrepancy} has been introduced by \cite{gelman1996posterior}, where $D(y, \theta)$ can be thought of as a generalization of a test statistic whose distribution can depend on $\theta$. 
\inlineRevised{For example, one may consider a generalization of the $\chi^2$ statistic \citep{gelman1996posterior}, defined as $\chi^2(y, \theta) = \sum_{i = 1}^n (y_i - \E(y_i \mid \theta ))^2/\V(y_i \mid \theta)$ or, more in general, the log-likelihood. The choice of the discrepancy typically depends on the context of the application.} 

Define $f(y^*, \theta | y) = p(y^* | \theta) p(\theta | y)$ as the joint posterior predictive distribution of $\theta$ and hypothetical data $y^*$ given observed data $y$.  The posterior predictive distribution of $y^*$ is $\int f(y^*, \theta | y) d\theta = f(y^* | y)$.
The $\ppp$ is the probability that the discrepancy measure calculated for a draw of the posterior predictive distribution is more extreme than the discrepancy measure of the data, averaged over the posterior $p(\theta | y)$. This is given by
\begin{equation}\label{eq:ppp_def_classic}
  \ppp(y) = \Pr \left\{D(Y^*, \Theta) \geq D(y, \Theta) | y\right\}.
\end{equation}
Letting   $\Delta(y^*, \theta | y) = D(y^*, \theta ) - D(y, \theta)$, we rewrite \eqref{eq:ppp_def_classic} as
\begin{equation}\label{eq:ppp_def}
  \ppp(y) =  \E_{(Y^*, \Theta)} \left[ \II{ \Delta(Y^*, \Theta | y ) \geq 0 } \right] , 
\end{equation}
where the expectation is over $(Y^*, \Theta) \sim f(y^*, \theta | y)$.  We interpret the $\ppp$ as a test statistic, or simply a statistic, rather than a ``p-value'', because it does not behave like a p-value. 

The $\cppp$ is designed to be the proper p-value for the $\ppp$ statistic.  That is, the $\cppp$ is defined as the probability, if the model is valid, of obtaining a $\ppp$ at least as extreme (small) as $\ppp(y)$.  In this case, the probability is over the space of hypothetical data, $\Yc \sim g(\yc | y)$, for some relevant reference density $g(\yc | y)$.  The $\cppp$ is given by
\begin{equation}\label{eq:cppp_def}
  \cppp(y) = \Pr\{\ppp(\Yc) \leq \ppp(y) | y\} = \E_{\Yc}\left[ \II{\ppp(\Yc) \leq \ppp(y)} \right].
\end{equation}
The distribution of $\ppp(\Yc)$, with $\Yc \sim g(\cdot)$, is used here as a null distribution for the test statistic $\ppp(y)$. The choice of the reference density $g(\cdot)$ is important.

The dependence of $g(\cdot)$ on the data $y$ means that even the $\cppp$ may not be perfectly calibrated.  If one knew the true parameters, $\theta_0$, then the choice $g(\yc | y) = g(\yc) =  p(\yc | \theta_0)$ would guarantee that $\cppp(y)$ is perfectly calibrated.  For example, it would give an accurate Type I error rate,  $\alpha$, if one chooses as a decision rule to reject the model if $\cppp(y) < \alpha$.  Since $\theta_0$ is unknown, in the examples below we choose the posterior predictive distribution, $g(\yc | y) = f(\yc, \theta | y)$.  As long as $\Pr\{\ppp(\Yc) \leq \ppp(y) | \theta \}$ does not vary strongly over values of $\theta$ with high posterior density, calibration based on this choice of $g(\cdot)$ should be fairly accurate.  This is similar to the issue that a parametric bootstrap may not be perfectly accurate because it depends on estimated parameters rather than true parameters.

Notice that different choices of the reference density $g(\cdot)$ lead to alternative definitions of Bayesian p-values. 
For example, \cite{hjort2006post} focus on the prior predictive distribution \citep{box1980sampling}, which does not depend on data $y$, with comments on other choices such as the posterior predictive. Other important variations and alternative Bayesian p-values are introduced in \cite{bayarri2000p} and further discussed by \cite{robins2000asymptotic} and \cite{bayarri2007bayesian}. 

Our choice is motivated by applied problems where the prior predictive distribution based on uninformative priors for $\theta$ would give a distribution of $\ppp(\Yc)$ very different from its distribution based on $\theta$ close to some $\theta_0$.  That is the distribution of $\ppp(\Yc)$, resulting from $\Yc \sim p(\yc | \theta)$, may vary substantially over very large ranges of $\theta$ such as the range supported by an uninformative prior.  

Note that the posterior predictive distribution has two roles.  It is the distribution of $(Y^*, \Theta)$ in (\ref{eq:ppp_def}), and it is a sensible choice for $g(\yc | y)$ in (\ref{eq:cppp_def}). These roles are distinct, as different choices of $g(\cdot)$ are possible. 

\section{Monte Carlo estimates of ppp and cppp}\label{sec:sec3}

\begin{figure}
\begin{center}
  \includegraphics[width=\textwidth]{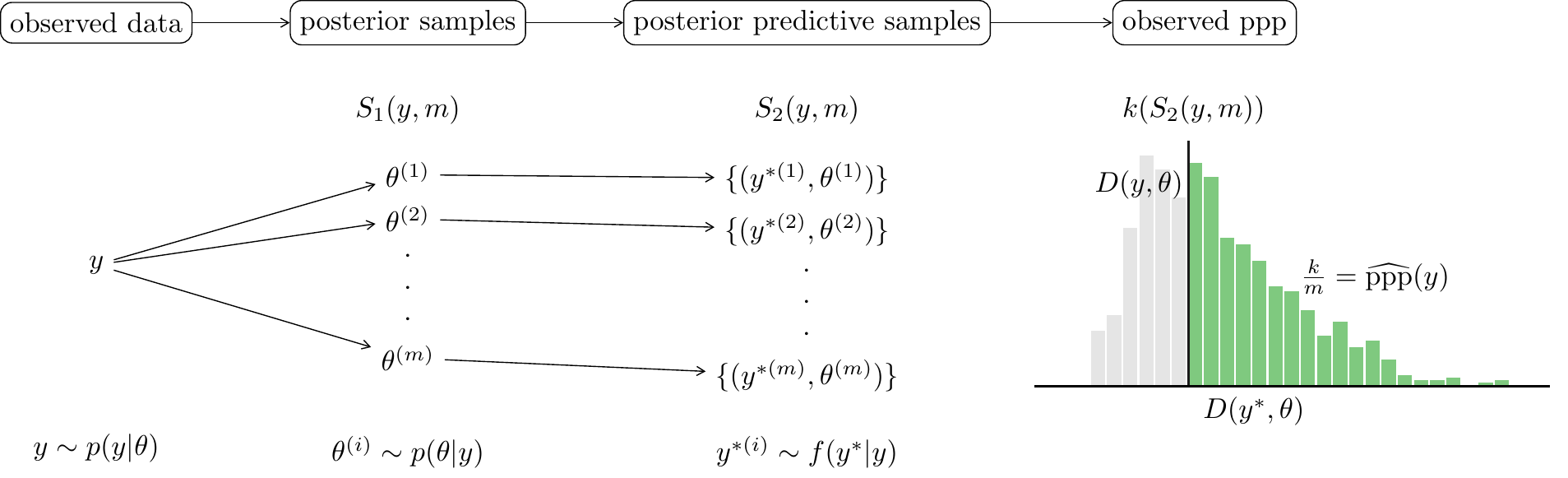}
  \caption{Schematic representation of the Monte Carlo approximation for the ppp.}
  \label{fig:ppp_monte_carlo}
\end{center}
\end{figure}

In practice, we work with Monte Carlo approximations of all expected values in \eqref{eq:ppp_def}-\eqref{eq:cppp_def}. Figure~\ref{fig:ppp_monte_carlo} provides a summary of the quantities involved in the estimation. Define a posterior sample of size $m$ based on the real data as $S_1(y, m) = \left\{ \trep \right\}, i = 1, \ldots, m$, resulting from an MCMC or other algorithm such that $\trep \sim p(\theta | y)$ for $i = 1, \ldots, m$. 
The $\trep$s will typically be sequentially correlated because they are sampled by MCMC. Define the collection of samples from the posterior predictive distribution based on the real data as $S_2(y, m)  = \left\{ (\yrep, \trep) \right\}, i = 1, \ldots, m$, with $\trep$ from $S_1(y, m)$ and $\yrep \sim p(y^* | \trep)$.  Define $k(S_2(y, m)) = \sum_{i = 1}^m \IIrep$, where $\IIrep = \II{\Delta(\yrep, \trep | y) \geq 0}$, so that $k$ is the count of posterior predictive discrepancies $D(\yrep, \trep)$ that are more extreme than data discrepancies $D(y, \trep)$. Then the standard Monte Carlo estimate of $\ppp(y)$ is
\begin{equation}\label{eq:ppp_hat}
  \ppph(y) = \frac{k}{m}.
\end{equation}
Notice that each $\IIrep$ draw is marginally distributed as Bernoulli with probability $\ppp(y)$. Hence, $k$ is a realization of a random variable parameterized by $y$ and $m$, and we can write $K \sim h(k | y, m)$.  If the posterior draws are independent, the distribution of $K$ is binomial, but in the general case, the $\IIrep$s are serially dependent. Using standard theory for MCMC output \citep[e.g.][Chapter 12]{robert1999monte}, define the integrated autocorrelation time of the chain of indicator variables $\left\{ \IIrep \right\}, i = 1, \ldots, m$, as $\tau$, so that its effective sample size is $\ESS = m/ \tau$. Then we have that the mean and variance of $K$ are
\begin{align*}
  \E[K | y] &= m \ppp(y),  \\
  \V[K | y] &= \frac{m^2 \ppp(y) (1-\ppp(y))}{\ESS}.
\end{align*}
For some purposes, one may want to replace $\frac{k}{m}$ with $\frac{k+0.5}{m+1}$, which is asymptotically equivalent but avoids zeros.

\begin{figure}
\begin{center}
  \includegraphics[width=\textwidth]{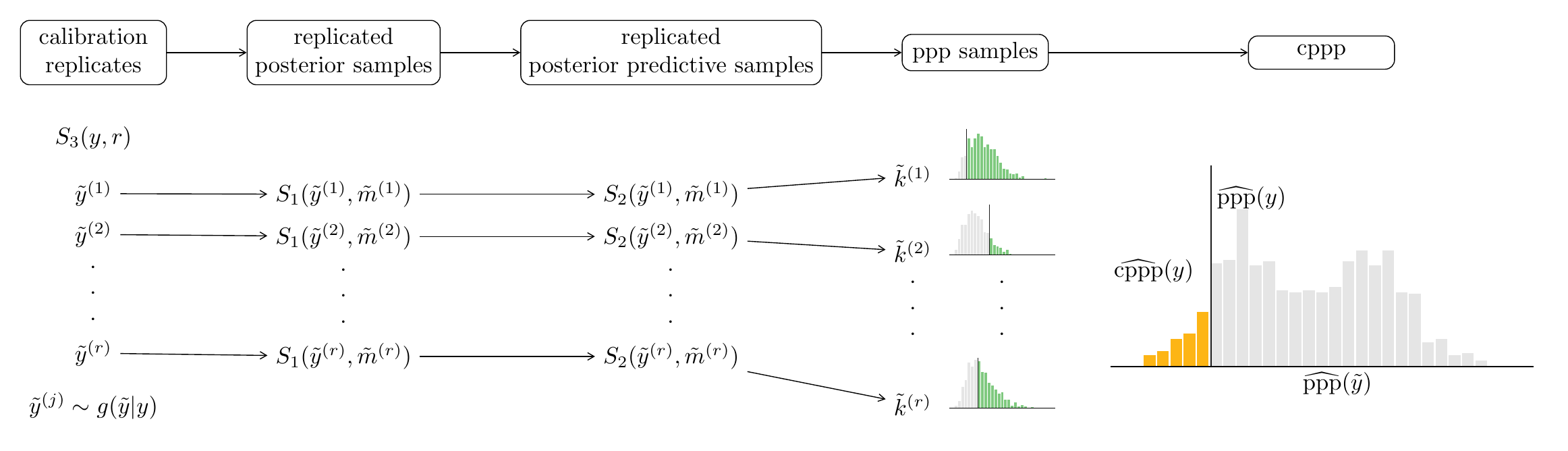}
  \caption{Schematic representation of the Monte Carlo approximation for the cppp.}  \label{fig:cppp_monte_carlo}
\end{center}
\end{figure}

The foregoing will be used once for the real data and then repeatedly for calibration replicates, as described next. For theoretical purposes, we assume that the number of MCMC samples $m$ is sufficiently large for the real data so that $\ppph(y)$ is very accurate, and hence assume that $\ppp(y)$ computed on observed data is essentially known, or at least has much smaller error than that of the calibration replicates.

To estimate $\cppp(y)$, we need a second layer of Monte Carlo approximation (see Figure~\ref{fig:cppp_monte_carlo}).  Define the collection of calibration replicates as $S_3(y, r) = \left\{ \ycrep \right\}$, $j = 1\ldots r$  with $\ycrep \sim g(\yc | y)$ and all $\ycrep$ mutually independent.  Given each $\ycrep$, new posterior samples $S_1( \ycrep, \mcrep )$ and samples from the posterior predictive $S_2( \ycrep, \mcrep)$ are drawn using a sample size choice, $\mcrep$, smaller from the $m$ used for the real data.  These new samples come from running the MCMC with the calibration replicate data, $\ycrep$, in the role of observed data. 
We initialize this MCMC with the same value of $\theta$ in $S_1(y,m)$ used to generate the calibration replicate data $\ycrep$. Hence we do not need a burn-in for the calibration replicates MCMC, since we are starting in the posterior region of interest.

From each sample, the value of interest is $\kcrep = k(S_2(\ycrep, \mcrep))$.  The Monte Carlo estimate of $\cppp(y)$ is then:
\begin{align}\label{eq:cppp_hat}
  \cppph(y) = \frac{1}{r} \sum_{j = 1}^r  \II{\kcrep \leq \mcrep \ppp(y)} = \frac{1}{r} \sum_{j = 1}^r  \II{\ppph(\ycrep) \leq \ppp(y) }.
\end{align}
It is important to notice that we are not interested in $\kcrep / \mcrep$ as an estimate of $\ppph(\ycrep)$. Rather, we care about it as an estimate of whether $\ppph(\ycrep)  \leq \ppp(y)$.  From here on we assume for simplicity that $\mcrep = \mc$ $\forall j$, i.e. all calibration replicates use the same sample size.

Since the reference density $g(\cdot)$ is chosen as the posterior predictive distribution based on the real data, $S_3(y,r)$ will in practice come from a subsample of $S_2(y,m)$.  Thus, $S_2(y, m)$ will be used twice: once to estimate the $\ppp$ of the real data, $\ppp(y)$, and once to approximate the distribution of $\ppp(\Yc)$, where $\Yc \sim g(\yc | y)$, for the purpose of estimating $\cppp(y)$. 
In the latter role, we aim to use relatively few samples to reduce computation, $r \ll m$, so the $S_3(y, r)$ samples are drawn randomly or by systematically thinning $S_2(y, m)$, making it reasonable to assume the $S_3(y, r)$ samples are mutually independent.

\subsection{Bias and variance of $\cppph(y)$}
\label{sec:bias-variance-monte}

To obtain insights on the role of the number of calibration replicates $r$ and MCMC samples $\mc$, we look at bias and variance of $\cppph(y)$ with respect to its theoretical definition.  The expected value of $\cppph(y)$ in (\ref{eq:cppp_hat}) is $\E_{\Yc} \left[\E_{\Kc|\Yc} \left[ \II{\Kc \leq \mc \ppp(y)} | \Yc \right]  \right]$, where $E_{\Yc}$ is expectation over $\Yc \sim g(\yc | y)$ and $E_{\Kc|\Yc}$ is expectation over $\Kc \sim h(\kc | \yc, m)$. The $\cppp(y)$ definition (\ref{eq:cppp_def}) can be written as $\E_{\Yc} \left[ \II{ \E_{\Kc|\Yc} \left[ \Kc | \Yc \right] \leq \mc \ppp(y) } \right]$.  The bias of $\cppph(y)$ is thus
\begin{equation}\label{eq:cppp_bias}
 \E_{\Yc} \left[\E_{\Kc|\Yc} \left[ \II{\Kc \leq \mc \ppp(y)}  | \Yc \right] - \II{\E_{\Kc|\Yc} \left[ \Kc | \Yc \right] \leq \mc \ppp(y) } \right].
\end{equation}
To gain some intuition about the bias, notice that the expression inside the outer expectation $\E_{\Yc}[\cdot ]$ is the average signed error of determining whether $\ppp(\yc)  \leq \ppp(y)$ using a draw $\Kc$, i.e. using the samples $S_2(\yc, \mc)$ (which would also give the estimate $\ppph(\yc)$).  When $\ppp(\yc)  \leq \ppp(y)$, the second indicator is 1, and $\E_{\Kc|\Yc} \left[ \II{\Kc \leq \mc \ppp(y)}  | \Yc \right]$ is the probability that $\ppph(\yc)  \leq \ppp(y)$ from a draw of $\Kc$.  When $\ppp(\yc)  > \ppp(y)$, the second indicator is 0, and $\E_{\Kc|\Yc} \left[ \II{\Kc \leq \mc \ppp(y)}  | \Yc \right]$ is one minus the probability that $\ppph(\yc)  > \ppp(y)$.  As $m \rightarrow \infty$, $\ppph(\yc) \rightarrow \ppp(\yc)$, and the bias goes to zero.

The variance of $\cppph(y)$ can be obtained using the law of total variance 
\begin{align}\label{eq:cppp_variance}
  \V[\cppph(y)] = & \frac{1}{r} \left\{ \E_{\Yc} \left[ \V_{\Kc | \Yc}(\II{\Kc \leq \mc \ppp(y)} | \Yc) \right] \right. \nonumber \\
  & + \left. \V_{\Yc} \left[ \E_{\Kc | \Yc} ( \II{\Kc \leq \mc \ppp(y)} |\Yc) \right] \right \}.
\end{align}
The first term is the average variance, due to Monte Carlo sampling of $\Kc$, of determining whether $\ppp(\yc)  \leq \ppp(y)$.  Hence this term will become small as $\mc$ gets large.  The second term is the variance across calibration samples $\yc$ of the probability that $\ppph(\yc)  \leq \ppp(y)$.  As $\mc$ gets large, this is the variance of a Bernoulli trial with success probability $\cppp(y)$, i.e. $\cppp(y) (1-\cppp(y))$.

\section{Computational costs and accuracy}\label{sec:sec4}

In this section, we study the costs and benefits of different computational allocations for the purpose of estimating $\cppph(y)$.  To do so, we skip the entire modeling exercise and assume that the null distribution of the $\ppp$ statistic is $\ppp(\Yc) \sim \mbox{Beta}(a, b)$, a Beta distribution with parameters $a > 0$, $b > 0$.  Although there are no actual $y$, $\yrep$ or $\ycrep$ values, we retain them in the notation to indicate the roles of different variables.  Under this scenario, $\cppp(y)$ is the tail area of the Beta distribution determined by the observed $\ppp(y)$: $\cppp(y) = F_{\mbox{Beta}}(\ppp(y), a, b)$, where $F_{\mbox{Beta}}(\cdot; a, b)$ indicates the cumulative density function of a $\mbox{Beta}(a, b)$ (see Figure~\ref{figSM0:example_cppp_scenario} for some examples). We also assume that independent draws can be made from the posterior and hence the posterior predictive distribution.

We are interested in how to allocate computational effort between calibration replicates ($r$) and posterior sample size for each calibration replicate ($\mc$).  We assume that posterior sampling is much more costly than any other step so that the total computational cost is $c \approx r \mc$.  For a given computational cost, we consider different allocations to $r$ and $\mc$.  Specifically, scenarios are created as follows:

\begin{enumerate}
\item \label{step:1} Choose $a$ and $b$ to set the null distribution of $\ppp(\Yc) \sim \mbox{Beta}(a, b)$.
\item Choose a value of $\cppp(y)$ to be estimated by (\ref{eq:cppp_hat}).
\item Set the corresponding observed $\ppp(y)$ as $\ppp(y) = F^{-1}_{\mbox{Beta}}(\cppp(y); a, b)$. We assume this is known via (\ref{eq:ppp_hat}) with large number of MCMC samples for the original chain $m$.
\item \label{step:4} Choose $c$, $r$ and $\mc$ such that $c = r \mc$.
\item For $j = 1\ldots r$:\label{step:iter-r}
  \begin{enumerate}
    \item Draw $\ppp(\ycrep) \sim \mbox{Beta(a, b)}$.  This is the unknown $\ppp$ for the $j$th calibration replicate.
    \item Draw $\kcrep \sim \mbox{Binomial}(\mc, \ppp(\ycrep))$.  This is the result of $\mc$ independent posterior samples (with data $\ycrep$) to estimate $\ppp(\ycrep)$ by $\ppph(\ycrep) = \kcrep / \mc$.
  \end{enumerate}
\item \label{step:cppph} Use the $r$ draws $\kcrep$ to calculate $\cppph(y)$ using (\ref{eq:cppp_hat}).
\end{enumerate}

\begin{figure}
  \includegraphics[width=\textwidth]{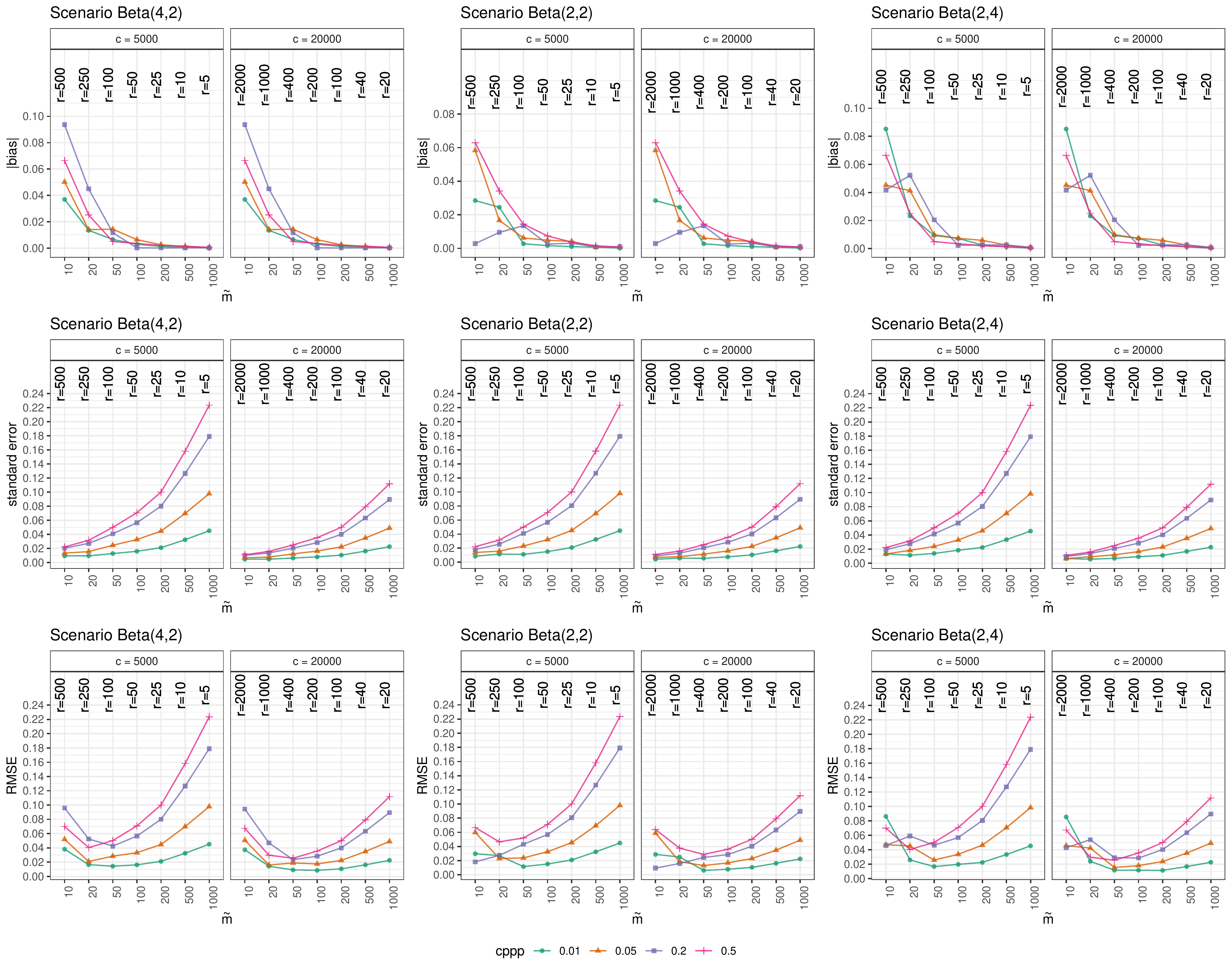}
  \caption{Absolute bias (top row), standard error (mid row) and RMSE (bottom row) for $\cppph(y)$ considering different values of $\cppp(y) \in \{0.01, 0.05, 0.20, 0.50\}$ and three different scenarios for the null distribution of $\ppp(\Yc)$ \inlineRevised{(a symmetric, right-, and left-skewed distribution).}}\label{fig2:results_beta_scenario} 
\end{figure}

This procedure aims at reconstructing the key quantities used in the Monte Carlo estimation of $\cppp(y)$ in a scenario where the null distribution of the ppp statistics is known, to compare estimation results at different computational costs. Since the marginal distribution of $\kcrep$ is Beta-Binomial with parameters $(\mc, a, b)$, we do not need to sample values, but we can compute the bias and variance of the $ \cppph(y)$ analytically.
The expected value of \eqref{eq:cppp_hat} corresponds to $F_{\mbox{Beta-Binomial}}(\mc \ppp(y); \mc, a, b)$, so that the bias with respect the true $\cppp(Y)$ is
\begin{equation*}
	F_{\mbox{Beta-Binomial}}(\mc \ppp(y); \mc, a, b) - F_{\mbox{Beta}}(\ppp(y); a, b),	
\end{equation*}
while the variance corresponds to
\begin{equation*}
	 \V[\cppph(y)] = \frac{1}{r} F_{\mbox{Beta-Binomial}}(\mc \ppp(y); \mc,  a, b)\left[ 1- F_{\mbox{Beta-Binomial}}(\mc \ppp(y); a, b)\right].
\end{equation*}
We calculate the bias and variance in \eqref{eq:cppp_bias}-\eqref{eq:cppp_variance} for different combinations of $(r, \mc)$ and choices of $\{a, b, \cppp(y)\}$ controlling the null $\ppp$ distribution. Some representative results are shown in Figure~\ref{fig2:results_beta_scenario}. We consider different values for the computational cost $c \in \{5,000; 20,000\}$. Values for the number of posterior samples $\mc$ of each calibration replicate are fixed to $\{10, 20, 50, 100, 200, 500, 1000\}$, with values for the number of calibration replicates $r$ defined correspondingly as $c/\mc$. Plots in Figure~\ref{fig2:results_beta_scenario} show absolute bias, standard error, and root-mean-squared error (RMSE = $\sqrt{\mbox{bias}^2 + \mbox{variance}}$) of $\cppph(y)$ under \inlineRevised{three} different scenarios for the null ppp distribution and four different values of $\cppp(y)$.

Results from these experiments suggest that values of $\mc$ in the range $50$--$200$ can be a good rule of thumb for choosing the number of posterior samples for each calibration replicate ($\mc$) in order to minimize the RMSE of $\cppph(y)$. The bias does not depend on the number of calibration replicates $r$, but it is controlled by $\mc$.  Very low values of $\mc$ can result in large bias, but even small values such as $50$--$200$ can make the bias negligible (see top row of Figure~\ref{fig2:results_beta_scenario}).  Then the RMSE is dominated by the variance, and it is better to allocate further computational effort to calibration replicates rather than posterior samples $\mc$. \inlineRevised{]. If the resulting confidence interval for the cppp is too large to reach a clear conclusion, more replicates can be added. In our empirical experiments, once the bias is negligible, we found that increasing replicates beyond $1,000$ would rarely be of interest (see Figure~\ref{figSM1:beta_scenario_fixedM} in the Supplementary Materials)}.

In the case of MCMC samples rather than independent samples, the values of $\mc$ in these results are to be interpreted roughly as the average ESS value of the chain $\{\IIrep\}, i = 1, \ldots, m$ across calibration replicates. Hence, depending on the mixing, one may need a larger number of MCMC samples to achieve a corresponding ESS of $50$--$200$. When the $\cppp(y)$ is not small, e.g., $0.20$ or $0.50$, even moderately large RMSE can be acceptable to determine whether a model will not be rejected. If $\mc$ is between $50$--$200$, then moderately large RMSE corresponds to a relatively small number of calibration replicates $r$, for example, $r = 100$.  Compare this to a naive implementation of $\cppp(y)$. That might use MCMC sample size of $m={10}^{3}$--${10}^{5}$ for the real data and the same MCMC sample size for each calibration replicate ($\mc = m$).  Most of the computational effort beyond $\mc$ = $50$--$200$ samples contributes little to the precision of $\cppph(y)$, and the total computational cost might be $1$--$3$ orders of magnitude lower by avoiding this naive approach.  However, for very small values of $\cppp(y)$, for example, close to common threshold values for statistical tests ($0.01, 0.05$), one may want to be conservative and choose higher values for $\mc$.

\section{Estimating the standard error of $\cppph(y)$}

\label{sec:sec5}
In practice, one needs to estimate whether $\cppph(y)$ has been approximated with sufficient precision to reach a clear interpretation.
If $\mc$ is sufficiently large to make the bias in \eqref{eq:cppp_bias} negligible, \inlineRevised{the RMSE of $\cppph(y)$ reduces to $\sqrt{\V[\cppph(y)]}$}. \inlineRevised{Then, an estimate of this quantity can be used to approximate Monte Carlo confidence intervals for the cppp, allowing inference on the model fit.}
In this section, we focus on how to estimate $\V[\cppph(y)]$ in \eqref{eq:cppp_variance} in practice, illustrating two possible estimators: plug-in and bootstrap.  In either case, the key challenge is to estimate $\V[\Kc | \Yc] = \V_{\Kc | \Yc}(\II{\Kc \leq \mc \ppp(y)} | \Yc)$ in the first term of \eqref{eq:cppp_variance} because it involves effective sample size with a potentially small MCMC sample.

\subsection{Plug-in estimator}

We obtain a plug-in estimator for the $\cppp$ variance by finding plug-in estimators for the two terms in \eqref{eq:cppp_variance}. The first term is the average variance, due to Monte Carlo sampling of $\Kc$, of determining whether $\ppp(\yc)  \leq \ppp(y)$.
Notice that, given $\Yc=\yc$, $\II{\Kc \leq \mc \ppp(y)}$ is a Bernoulli random variable with probability of success $\Pr(\Kc \leq \mc \ppp(y)) = F_{\Kc}(\mc \ppp(y))$ where $F_{\Kc}(\cdot)$ is the cumulative density function of $\Kc$. Then the variance of this indicator variable can be expressed as:
\begin{equation*}
  \V_{\Kc | \yc}\left(\II{\Kc \leq \mc \ppp(y)} | \Yc\right) = F_{\Kc}(\mc \ppp(y))  \left[1 - F_{\Kc}(\mc \ppp(y) )\right].
\end{equation*}
As noted in Section~\ref{sec:sec3}, $\Kc$ is a random variable with mean $\mc \ppp(\yc)$ and variance $\tauc \mc \ppp(\yc)(1 - \ppp(\yc))$, where $\tauc$ is the integrated autocorrelation time for an MCMC chain using data $\yc$.  Hence $\Kc$ is like a binomial random variable with variance inflated by $\tauc$.  We use a normal approximation with a continuity correction for the distribution of $\Kc$, so that
\begin{equation}\label{eq:normalApproxKc}
  F_{\Kc}(\mc \ppp(y) ) \approx \hat{F}_{\Kc}(\mc \ppp(y)) = F_{\mbox{N}}\left(\mc \ppp(y) + \frac{1}{2}; \mc \ppp(\yc), \tauc \mc \ppp(\yc)\big(1 - \ppp(\yc)\big) \right)
\end{equation}
where $F_{\mbox{N}}(\cdot; \mu, \sigma^2)$ denotes the cumulative distribution function of a Normal distribution with mean $\mu$ and variance $\sigma^2$.
In practice, we can use $\ppph (\ycrep)$ as a plug-in estimate for $\ppp(\yc)$ for each calibration replicate $j = 1, \ldots, r$ so that the first term of $\eqref{eq:cppp_variance}$ is approximated via Monte Carlo as
\begin{equation*}
 \E_{\Yc}\left[\V[\Kc | \Yc]\right] \approx \frac{1}{r} \sum_{j = 1}^r \hat{F}_{\Kc}^{(j)} \left[1 - \hat{F}_{\Kc}^{(j)}\right]
\end{equation*}
where $\hat{F}_{\Kc}^{(j)} = F_{\mbox{N}}\left(\mc \ppp(y) + \frac{1}{2}; \mc \ppph(\ycrep), \hat{\tauc}^{(j)}\mc \ppph(\ycrep)(1 - \ppph(\ycrep)) \right)$. The challenge here is to estimate the autocorrelation time $\tauc^{(j)}$ for each calibration replicate, as we would like to use a small number of MCMC samples; we propose an estimator in \ref{sec:subsec_transfer}.

For the second term in $\eqref{eq:cppp_variance}$, we have that the inner expectation, given $\Yc=\yc$, is
\begin{equation*}
\E_{\Kc | \yc} \left( \II{\Kc \leq \mc \ppp(y)} | \yc\right) = \Pr(\ppph(\yc) \leq \ppp(y)) = F_{\Kc}(\mc \ppp(y)).
\end{equation*}
Note that for large $\mc$ and/or $\ppp(\yc)$ far from $\ppp(y)$, $F_{\Kc}(\mc \ppp(y)) \approx \II{\ppp(\yc) \leq \ppp(y)}$.

Using the normal approximation for $F_{\Kc}(\cdot)$ and Monte Carlo approximation of the outer variance, and combining the two terms of (\ref{eq:cppp_variance}) gives the final plug-in estimator as

\begin{align}\label{eq:var_estimate}
  \widehat{\V}[\cppph(y)] &=  \frac{1}{r} \sum_{j = 1}^r \hat{F}_{\Kc}^{(j)} \left[1 - \hat{F}_{\Kc}^{(j)}\right] + \frac{1}{r} \sum_{j = 1}^r \left[ \left(\hat{F}_{\Kc}^{(j)} - \overline{F_{\Kc}} \right)^2 \right] \nonumber \\
  &= \overline{F_{\Kc}}(1-\overline{F_{\Kc}})
\end{align}
where $\overline{F_{\Kc}} = \sum_{j = 1}^r \hat{F}_{\Kc}^{(j)}/r$.  Notice that, as $\mc \rightarrow \infty$: $\hat{F}^{(j)}_{\Kc} \rightarrow \ppp(\yc)$ and $\overline{F_{\Kc}} \rightarrow \cppph(y)$.

\subsection{Bootstrap estimators}

A bootstrap procedure can be used to estimate the variance of $\cppph(y)$. The main idea is to obtain $b$ bootstrap estimates $\{ \cppph(y)^{(l)} \}_{l = 1, \ldots, b}$, and evaluate the sample variance of these estimates. To do so we resample the calibration replicates and MCMC samples used for the original $\cppp(y)$ estimate, taking into account the structure of the Monte Carlo estimation procedure in Figures~\ref{fig:ppp_monte_carlo}-\ref{fig:cppp_monte_carlo}.

The idea is the same as bootstrapping hierarchical data, where observations are associated with groups, for example, students belonging to schools. A bootstrap of this kind of data is typically performed in two stages to retain the hierarchical structure of the data: first groups are resampled, then observations are resampled within the groups. In our case, we first resample the calibration replicates, and then the MCMC samples within each replicate. In practice, instead of resampling all the MCMC outputs, we resample only the derived output $\left\{ \Delta(\yrep, \trep | \yc) \right\}, i = 1, \ldots, \mc$, for a calibration replicate $\yc$.  Only these derived quantities are needed to obtain the count $\kc$. In this step we need to retain serial dependence between the MCMC samples, so we can either use moving block bootstrap (MBB) \citep{mignani1995mbb} or exploit the normal approximation for $\Kc$.

In the following, we summarize the steps to obtain these two bootstrap estimates. For a fixed number of bootstrap samples $b$, each $\cppph(y)^{(l)}$, for $l = 1, \ldots, b$, is obtained as follows:
\begin{enumerate}
  \item Sample with replacement $r$ calibration replicates from $S_3(y,r)$. We denote with $S_3^{(l)}(y, r)$ the $l$-th  bootstrap sample.
  \item For each calibration replicate $j$ in $S_3^{(l)}(y, r)$ perform one of the following:
  \begin{itemize}
    \item[a)] Bootstrap-MBB: Calculate $\{\kcrep\}$ from a moving block bootstrap (re)sample of the chain $\left\{ \Delta(\yrep, \trep | \ycrep) \right\}, i = 1, \ldots, m$.
          Then use counts $\{\kcrep\}^{(l)}$ to obtain $\cppph(y)^{(l)}$ as in \eqref{eq:cppp_hat}.
    \item[b)] Bootstrap-normal: draw $\{\kcrep\}$ from the normal approximation with mean and variance as stated in (\ref{eq:normalApproxKc}).
  \end{itemize}
  \item Use counts $\{\kcrep\}^{(l)}$, $j=1,\ldots,r$, to obtain $\cppph(y)^{(l)}$ as in \eqref{eq:cppp_hat}.
\end{enumerate}
Although the moving block bootstrap accounts for the original correlation structure between MCMC samples, the number and sizes of blocks are additional parameters to choose in comparison with the normal approximation. Nevertheless, in both cases, one has to choose the number of replicates which adds to the $\cppp(y)$ computational cost. 

\subsection{A transfer estimator for the integrated autocorrelation time}\label{sec:subsec_transfer}

For the normal approximation used in the first term of the plug-in estimate and the boostrap-normal approach, we need an estimate of the integrated autocorrelation time $\tauc$, or the corresponding $\ESSc  = \mc / \tauc$ for the Markov chain $\left\{ \IIci \right\}$, $i = 1 \ldots \mc$. The index $(j)$ for the calibration replicate will be omitted in this section.

Each $\IIci$ is an indicator value from the posterior predictive sample for calibration replicate $\yc$.  That is, $\IIci = \II{\Delta(\yrep, \trep | \yc) \geq 0}$, where $S_2(\yc, \mc) = \left\{ (\yrep, \trep) \right\} i = 1, \ldots, \mc$ is the MCMC output from sampling $(\yrep, \trep) \sim p(y, \theta | \yc)$.  However, methods for estimating $\tauc$ or  $\ESSc$ typically need long chains, whereas here we seek to reduce computation by using relatively short chains for calibration replicates.

On heuristic grounds, we propose to use information from the single long MCMC chain for the real data to provide reasonable estimates of $\tauc$ for the calibration replicates. Specifically, we assume that the autocorrelation structure of the chain $\left\{ \Delta(\yrep, \trep | y) \right\}, i = 1, \ldots, m$ will be similar for different values of $y$.  This is the chain of differences between posterior predictive discrepancies and data discrepancies.  It is important to use this chain rather than the chain of indicators, $\left\{ \IIci \right\}$, because the correlation structure of the latter will clearly depend on $\yc$.  It is a chain of 0s and 1s with probability $\ppp(\yc)$ of being 1, so for some values of $\yc$ the $\IIci$ may be mostly 0s, for others mostly 1s, and so on.  To reflect this dependence, we write $\tauc$ as $\tauc(\ppp(\yc))$.

The strategy will be as follows.  For an estimated $\ppp$ for a calibration replicate, $q = \ppph(\yc)$, we will use the shifted chain $\left\{ \Delta(\yrep, \trep | y) - \Delta_{q} \right\}$,  where $\Delta_q$ is the $q$ quantile of $\left\{ \Delta(\yrep, \trep | y) \right\}$.  This satisfies $\frac{1}{m} \sum \II { \Delta(\yrep, \trep | y) - \Delta_q \le 0 } = q$ and thus corresponds to a chain that yields an estimated $\ppp$ equal to that of the calibration replicate.

Then, if the above assumption is reasonable, the indicators of the shifted chain,
\begin{equation}\label{eq:shifted_Ind}
\II { \Delta(\yrep, \trep | y) \le \Delta_{q} },
\end{equation}
should have integrated autocorrelation time close to that of $\tauc(q)$, i.e. of chain of indicators for the calibration replicate.  The $\ESS$ (and hence $\tauc$) for the chain in (\ref{eq:shifted_Ind}) can be estimated by any suitable method for a discrete Markov chain such as the batch mean estimator of \cite{flegal2010batch}, implemented in the \texttt{mcmcse} package \citep{flegal2020MCMCse}.

\section{Examples}\label{sec:sec6}

In this section, we consider two examples that have been used in the literature \citep{hjort2006post}. The first example uses a normal model for data with outliers, while the second comprises two versions of a classic example of a capture-recapture model. To estimate the models used in each example we make use of the NIMBLE software \citep{nimble-software:2022} a flexible R-based system for hierarchical models. The procedure for cppp is implemented using NIMBLE's algorithm programming system and can be used with other models and discrepancies provided by the user.

These examples are difficult because the $\cppp$ for all cases are near or below the traditional rejection threshold of 0.05. Since the most dramatic computational gains of our proposed methods will be relevant when the $\cppp$ is well above such a threshold (when a model is acceptable) we also include a version of the capture-recapture example with simulated data.
We use these examples to i) assess how results from the simulated experiments in Section~\ref{sec:sec4} hold in real-data applications and ii) compare the different estimators for the $\cppp$ standard error illustrated in Section~\ref{sec:sec5}. To do so, we look at $\cppph(y)$ and its standard error under different scenarios, considering the computational cost, its allocation and mixing properties of the original MCMC run.  We show averages of $\cppph(y)$ and each kind of standard error estimate from $500$ runs of the whole procedure (Figures~\ref{fig:newcomb_results}-\ref{fig:simulated_cr}). In the Supplementary Materials, we also report \inlineRevised{times (Table~\ref{tab:times})}, coverage for approximate confidence intervals at 95\% (Table~\ref{tab:coverage}) and estimates of the ppp distribution for each example (Figure~\ref{fig:ppp_distr}).

Since we are working with real data examples, we do not have a ground truth to compare with. Hence, we use brute force computing to obtain good estimates of the truth for comparison. We obtain brute force estimates of $\cppp$ by choosing large $r$ and $\mc$. We obtain brute force Monte Carlo estimates of the standard error of $\cppp$ as the standard deviation of $\cppph(y)$ from $500$ repeats of the entire $\cppp$ computation (illustrated in Figure~\ref{fig:cppp_monte_carlo}), for each scenario of $r$ and $\mc$ considered.

\subsection{Normal model example using newcomb data} 

A standard example to illustrate posterior predictive p-values in the literature uses the speed of light data \cite[see for example][Section 6.3]{gelman2013bayesian}, that comprises $66$ measurements of the speed of light made by Simon Newcomb in $1882$. This is a classic example of a dataset with outliers.
A standard model for the data would use a normal distribution with unknown mean $\mu$ and variance $\sigma^2$, however, there are two extreme low measurements not compatible with the model. As in \cite{gelman2013bayesian}, we consider as discrepancy measure $D(y, \mu) = |y_{(61)} - \mu| - |y_{(6)} - \mu|$, where $y_{(i)}$ denotes the $i$th-ordered data point. Assuming a uniform prior for $(\mu, \log\sigma)$, \cite{gelman2013bayesian} report $\ppph(y) = 0.26$ using $200$ simulations from the posterior predictive. We replicated the analysis considering the same model, using $m =4,000$ samples after a burn-in of $1,000$ samples, obtaining $\ppph(y) = 0.205$. This result is in line with \cite{hjort2006post}, who reported a value of $0.208$ using 1 million samples. Although the authors did not compute a value for the $\cppph(y)$, they speculate that it falls on ``the statistical borderline of surprise''\inlineRevised{, assuming a threshold of $0.05$ for statistical significance.} We in fact estimate $\cppph(y) = 0.055$ using $r = \mc = 1,000$. 
\begin{figure}
\begin{center}
  \includegraphics[width=\textwidth]{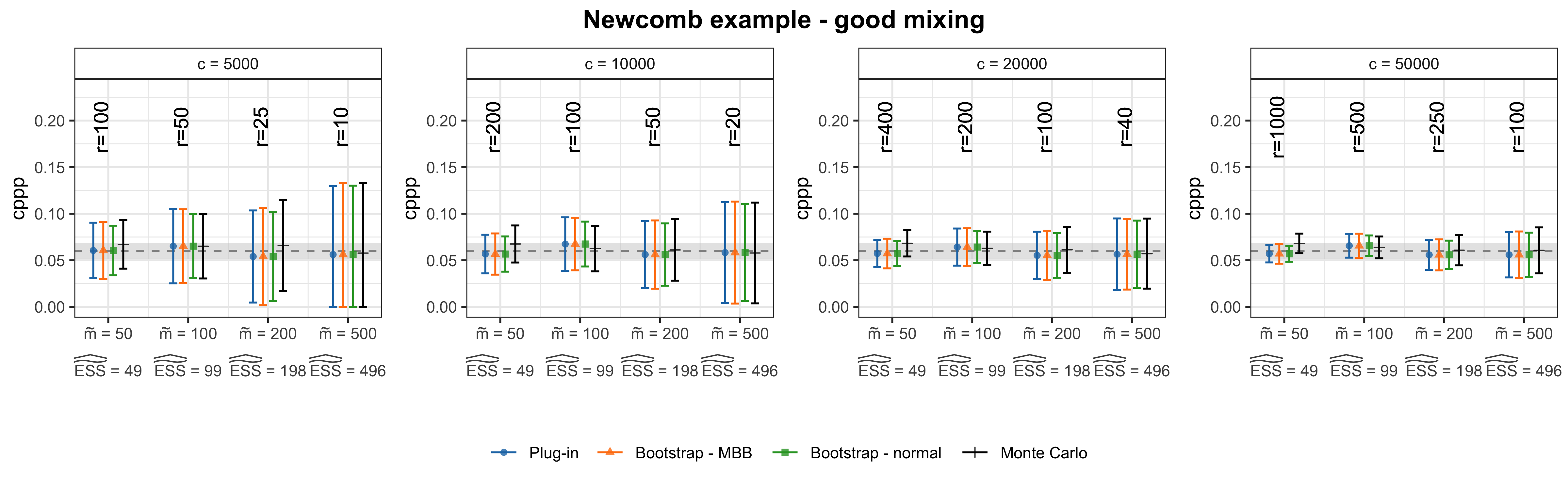}
  \includegraphics[width=\textwidth]{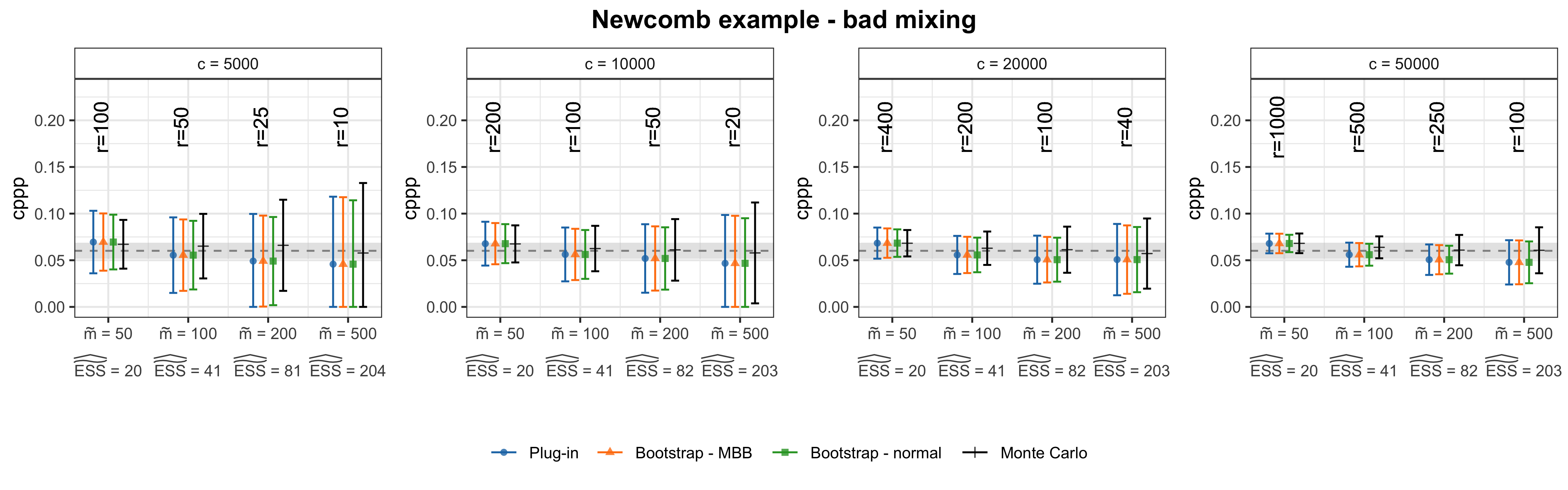}
  \caption{Comparison of estimates of $\cppph(y)$ and its standard error for different combinations of calibration replicates $r$ and MCMC samples $\mc$ for fixed computational cost $c$, under good mixing conditions (top row) and bad ones (bottom row). Error bars correspond to one standard deviation, estimated using different methods. The ``Monte Carlo'' (black) case shows brute force estimate of average $\cppph(y)$ and its standard error.
  The dashed gray line shows brute force estimate of the correct $\cppp(y)$
  while the shaded gray area is its Monte Carlo standard deviation.}
  \label{fig:newcomb_results}
\end{center}
\end{figure}
Similarly to the experiments in Section~\ref{sec:sec4}, we consider $4$ values for the total computational cost $c \in \{5,000; 10,000; 20,000; 50,000\}$, fix values of $\mc \in \{50, 100, 200, 500\}$, and define $r = c/\mc$. We also consider two different conditions for the mixing of the original MCMC. In the first case, we have fairly good mixing for the model parameters, where $\ESS/m$ for $(\sigma, \mu)$ is $(0.2, 1)$; in the second one, we used random walk Metropolis-Hastings samplers for posterior sampling and purposely tune them to induce bad mixing, so that the resulting $\ESS$ is 4 to 6 times lower than in the first case, i.e. $\ESS/m$ for $(\sigma, \mu)$ is (0.05, 0.15).

Figure~\ref{fig:newcomb_results} shows averages (from $500$ runs) of $\cppph(y)$ and of each standard error estimator, for different mixing conditions (subfigure rows), fixed computational costs (subfigures), and combinations of calibration replicates and MCMC samples ($r$ and $\mc$ values). Error bars represent one standard deviation. The colored error bars show the different estimators described in Section~\ref{sec:sec5}, while the black bar denotes the brute force Monte Carlo estimate. Results for the bootstrap estimators are computed using $100$ replications. We also report, for each value of $\mc$, the average ESS for the chain $\{\IIrep\}, i = 1, \ldots, m$ across calibration replicates, denoted as $\widehat{\widetilde{\mbox{ESS}}}$. We use the transfer method described in Section~\ref{sec:subsec_transfer} to estimate the ESS for each calibration replicate and report the mean value averaged across calibration replicates and the $500$ runs. 

Under the good mixing conditions, we obtain comparable results to the brute force values for the $\cppp(y)$ estimate; in the bad mixing case, we observe small but clear bias when either the number of MCMC samples or calibration replicates is small (e.g., $\mc = 50$ corresponding to $\mbox{ESS} = 20$ or $r \leq 50$). The different variance estimators seem to perform similarly under both mixing conditions. However, there are relevant differences between the estimators in terms of coverage (see Table~\ref{tab:coverage}). Intervals based on the Plug-in or Bootstrap-MBB estimators are close to the nominal 95\% level when there are good compromises between ESS and the number of calibration replicates ($\mbox{ESS} \geq 50$ and $r\geq100$). Instead, the bootstrap normal estimator leads more often to under-coverage, which becomes more evident under bad mixing conditions. In settings with low ESS values and a high number of calibration replicates, the coverage can be 0 due to biased cppp estimates with low variance, which results in confidence intervals that do not cover the true cppp.

\subsection{Capture-recapture models using dipper data}

Another example from the literature \citep{hjort2006post,nott2018approximation}, considers capture-recapture models using data on the European Dipper (\textit{Cinclus cinclus}). Data consists of $n= 294$ sighting histories collected over $k= 7$ annual sighting occasions from $1980$ to $1987$. A simple capture-recapture model can be described as a Hidden Markov Model, where each sighting is a realization of a Bernoulli random variable $Y_{i, t} \in \{1 = \mbox{seen}, 0 = \mbox{not seen}\}$, conditional to a binary latent variable $X_{i,t}\in \{1 = \mbox{alive}, 0 = \mbox{dead}\}$. The model is parametrized by the annual probability of survival $\boldsymbol{\phi} = (\phi_1, \ldots, \phi_{k -1})$ and the probability of being seen $\boldsymbol{p} = (p_2, \ldots, p_k)$. The hierarchical model corresponds to:
\begin{align*} 
  X_{i,t} | X_{i,t -1}  &\sim \mbox{Bernoulli}(\phi_{t -1} x_{i,t-1}), \quad t = 2, \ldots, k, \\
  Y_{i,t} | X_{i,t}     &\sim \mbox{Bernoulli}(p_t x_{i, t}) \quad t = 2, \ldots, k\\
  \phi_{t-1} & \sim \mbox{Uniform}(0,1) \quad  p_t \sim \mbox{Uniform}(0,1), \quad t = 2, \ldots, k,
\end{align*}
where $X_{i,1} = Y_{i,1} = 1$, for $i = 1, \ldots, n$. A simpler version of the model above assumes constant survival and capture probabilities over time, so that $p_2 = \ldots = p_k = p$ and  $\phi_1 = \ldots = \phi_{k-1} = \phi$. Under both models, uniform and independent priors are chosen for the survival and capture probabilities. \cite{hjort2006post} considers these two versions of the model, following \cite{brooks2000bayesian}, referring to those as the \emph{large model} (T/T model) and the \emph{small model} (C/C model) respectively.

To evaluate whether the models characterize the data well,  \cite{brooks2000bayesian} use the Freeman-Tukey statistic \citep{freemen1954transformations} as a discrepancy measure 
\begin{equation}
	D(y, \theta) = \sum_{s = 1}^{k -1} \sum_{t= 2}^{k} (\sqrt{z_{st}} - \sqrt{e_{st}})^2,
\end{equation}
where $z_{st}$ is the observed number of animals released at time $s$ and captured at time $t$, while $e_{st}$ is the corresponding expected number.

For the original MCMC run we follow \cite{brooks2000bayesian} and consider $10,000$ samples after $1,000$ burn-in. We obtain values for the observed ppp of $0.064$ for the C/C model, and $0.083$ for the T/T, slightly differently from \cite{brooks2000bayesian} who reported $0.069$ and $0.086$ respectively. However, we can attribute this small difference to Monte Carlo variation. \cite{hjort2006post} reports different values, $0.060$ for the C/C model, and $0.075$ for the T/T model; we obtain the same values when considering only $1,000$ samples after burn-in. To compute the cppp, \cite{hjort2006post} use $500$ calibration replicates without specifying the number of MCMC samples, obtaining cppp values of $0.022$ and $0.002$ under the two models respectively. We obtain slightly different values. Considering a naive estimate using $r = 1,000$ and $m = 10,000$, the cppp is equal to $0.044$ for the C/C model and $0.01$ for the $T/T$ model, similar to the Monte Carlo baseline. \inlineRevised{However, we reach the same conclusion as in \cite{hjort2006post}, that the cppps reject the models, whereas the ppps do not, and the ordering of which model deserves more skepticism is flipped.}
%
\begin{figure}
\begin{center}
  \includegraphics[width=\textwidth]{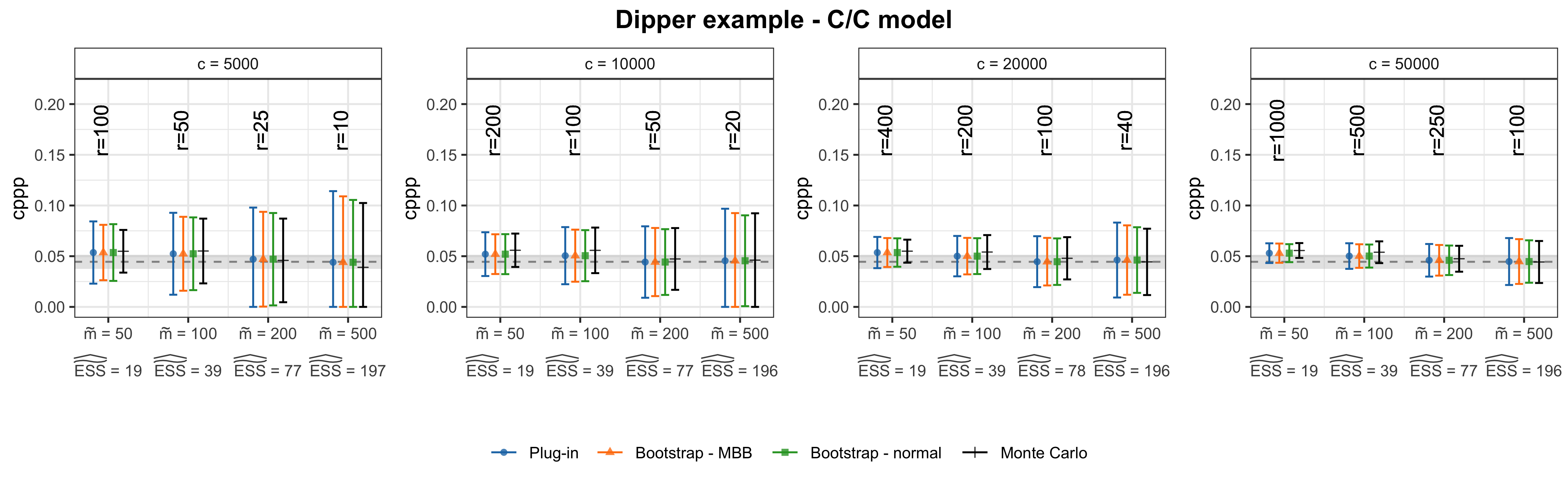}
  \includegraphics[width=\textwidth]{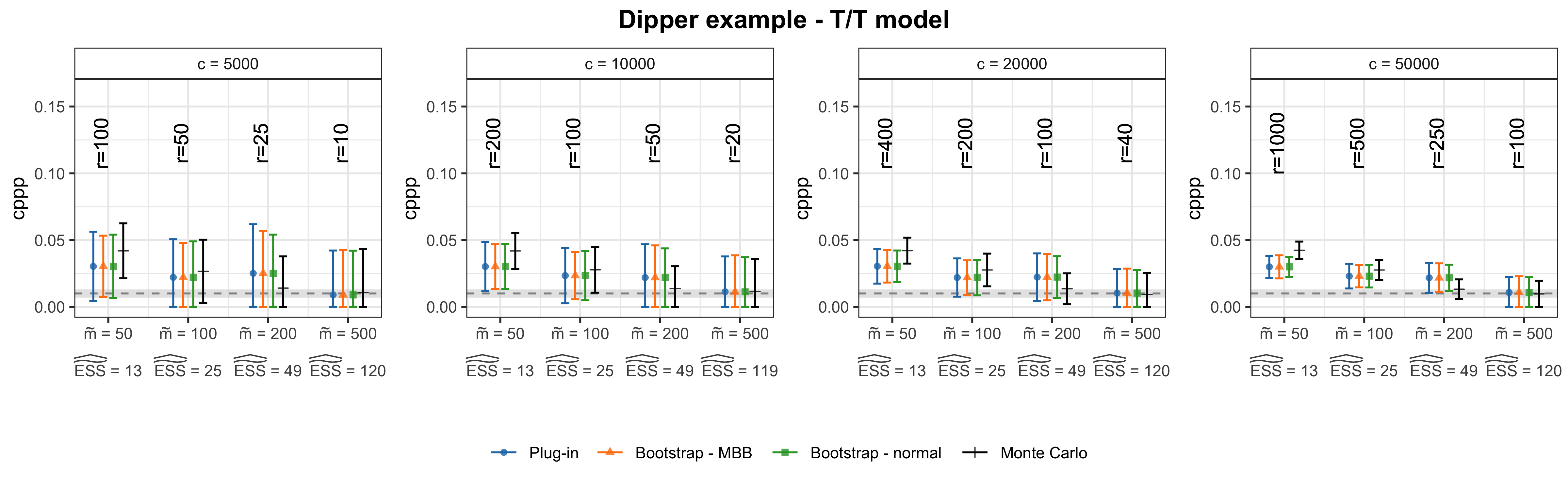}
  \caption{Comparison of estimates of $\cppph(y)$ and its standard error for different combinations of calibration replicates $r$ and MCMC samples $\mc$ for fixed computational cost $c$ under the C/C model (top row) and T/T model (bottom row). Error bars correspond to one standard deviation, estimated using different methods. The ``Monte Carlo'' (black) case shows brute force estimate of average $\cppph(y)$ and its standard error.
  The dashed gray line shows brute force estimate of the correct $\cppp(y)$
  while the shaded gray area is its Monte Carlo standard deviation.}
  \label{fig:dipper_comparison}
\end{center}
\end{figure}
As in the previous example, we compare the performance of the standard deviation estimators for the two models (Figure~\ref{fig:dipper_comparison}). Since for both models we obtain significant cppps, i.e. values under the commonly used threshold of $0.05$, we also simulated a case where the cppp is not significant, generating data under the T/T model (Figure~\ref{fig:simulated_cr}).  This is important because cases where the $\cppp$ turns out to be acceptable are the ones where the most computation can be saved since rough estimates of $\cppp$ may be satisfactory.

For both models, our approximate results are comparable to the brute force values under some combinations of $m$ and $r$. However, we obtain biased point estimates of the $\cppp(y)$ when either the number of calibration replicates or independent MCMC samples is low. In particular, for $r < 50$ we have biased estimates under the C/C model, while for the T/T model we obtain a good point estimate when the ESS is at least $100$ ($\mc = 500$) and $r>50$.  This behavior can be related to the different shapes of the null $\ppp(Y)$ distributions under the two models (Figure~\ref{fig:ppp_distr}). Under the C/C model, the ppp statistic has a symmetric and almost flat distribution. However, for the T/T model we observe a left-skewed distribution, similarly to the two examples in Section~\ref{sec:sec4}. For similar values of the ESS under the two models, we need a different number of calibration replicates to accurately estimate the $\cppp$. This is because when the null distribution of $\ppp(Y)$ is skewed and the $\cppp(y)$ is very small, it is hard to sample values of ppp smaller than the observed $\ppp(y)$, (i.e. in the tail of the distribution); hence a higher number of calibration replicates is needed to achieve the same accuracy.

Considering the $\cppp$ standard error, we observe good performances of all estimators at high computational cost ($c = 50,000$). The full nonparametric bootstrap (bootstrap-MBB) is more robust to low ESS values than the bootstrap-normal, at the cost of adding more steps in the computation. The plug-in estimator performs quite well overall, making it the preferable option. The same considerations apply to the performance of the intervals in terms of coverage (Table~\ref{tab:coverage}).
\begin{figure}
\begin{center}
  \includegraphics[width=\textwidth]{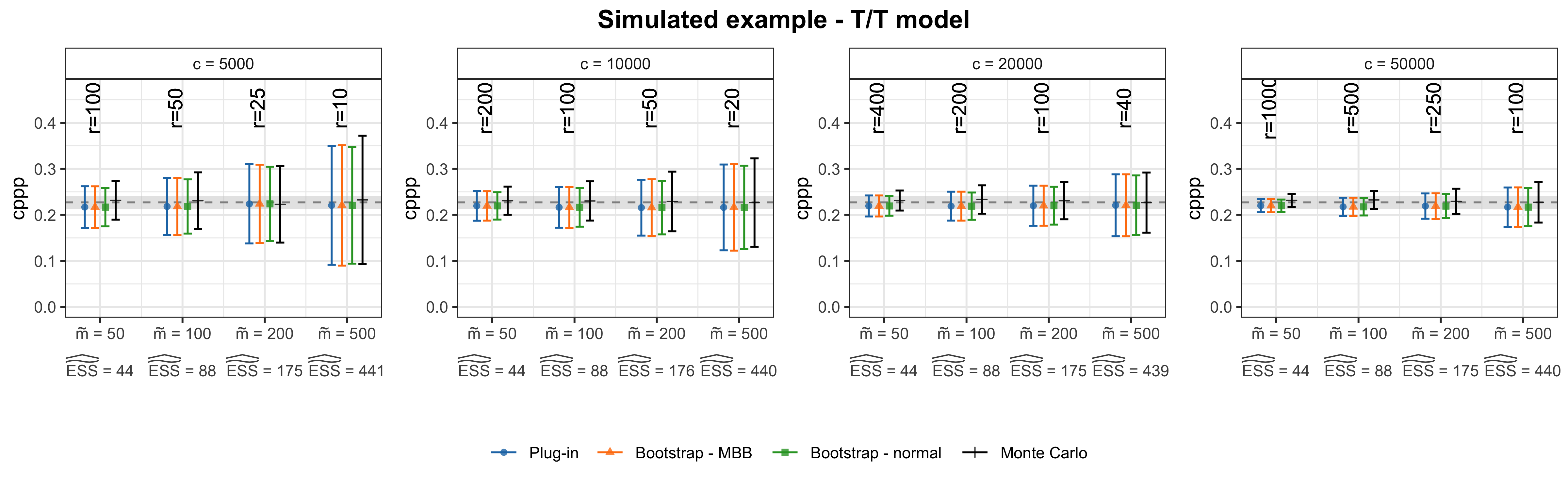}
  \caption{Comparison of estimates of $\cppph(y)$ and its standard error for different combinations of calibration replicates $r$ and MCMC samples $\mc$ for fixed computational cost $c$ for a simulated example using the T/T model. Error bars correspond to one standard deviation, estimated using different methods. The ``Monte Carlo'' (black) case shows brute force estimate of average $\cppph(y)$ and its standard error. The dashed gray line shows brute force estimate of the correct $\cppp(y)$ while the shaded gray area is its Monte Carlo standard deviation.}
  \label{fig:simulated_cr}
\end{center}
\end{figure}
Finally, in Figure~\ref{fig:simulated_cr} we report results for data simulated from the T/T model, for which the brute force $\cppp$ is $0.23$. In this case, even results from the lowest computational cost ($c =5,000$) are accurate enough to conclude that the model will not be rejected. One hopes that models are often well chosen and hence this substantial reduction in computational cost to conclude a model is acceptable can be realized in many applications. 

\section{Discussion}

This paper gives computational methods to efficiently assess the goodness of fit of Bayesian models via calibrated posterior predictive p-values. Our proposal allows obtaining the cppp much faster than via naive implementation, i.e., when the calibration procedure uses the same number of iterations as in the original MCMC.
The main takeaway is that good choices for the number of calibration replicates and MCMC samples per replicate can reduce the computational burden by orders of magnitude in model assessment without significant loss in accuracy. These two quantities control the bias and variance trade-off of the cppp estimate.
\inlineRevised{A good rule of thumb for the number of MCMC samples is using a number of iterations such that the ESS of the indicator used in the ppp calculation, is within $50$--$200$, making the bias negligible. This can be in practice checked during the MCMC sampling and implemented as a stopping rule. As default we suggest starting with $50$--$100$ calibration replicates and use an estimate of the cppp variance to inform whether more replications are needed, which can be readily parallelized.} \inlineRevised{In practice, the most computation is demanded when the cppp is close to a decision threshold, while the greatest savings occur when model adequacy can be confirmed using a rough estimate of the $\cppp$  with accurately quantified uncertainty.}

In this paper, we also illustrate different methods to quantify the uncertainty associated with the cppp estimate, which rely on plug-in estimators or bootstrap procedures. We find that the plug-in estimate for the cppp standard error performs generally well in terms of accuracy, and is preferred because it does not add computational effort as do the bootstrap procedures. This is important in that accurate and fast uncertainty quantification for the cppp estimate allows for immediate improvements in the procedure. When the cppp is not small, even a rough estimate of the cppp obtained at low computational cost, is enough to conclude that the model will not be rejected. 

The proposed procedure to approximate the cppp is implemented using the NIMBLE software, which is a flexible platform that can handle a wide range of models and discrepancies. This software provides users with the tools to efficiently implement our proposed procedure, allowing the use of the cppp methodology beyond toy examples.

\newpage
\begin{center}
{\large\bf SUPPLEMENTARY MATERIAL}
\end{center}

\begin{description}
\item[fastCPPP:] folder containing the R scripts to reproduce plots in Section 4, results for the examples in Section 6, and results in the Supplementary Materials.\textbf{}
\item[Supplementary figures and tables:] A .pdf file containing supplementary figures and tables. 
\end{description}

\newpage

\section*{Supplementary figures and tables}

\renewcommand{\thefigure}{S\arabic{figure}}
\setcounter{figure}{0}
\renewcommand{\thetable}{S\arabic{table}}
\setcounter{table}{0}

\subsection*{Supplement to Section~3}

\begin{figure}[h]
  \includegraphics[width=\textwidth]{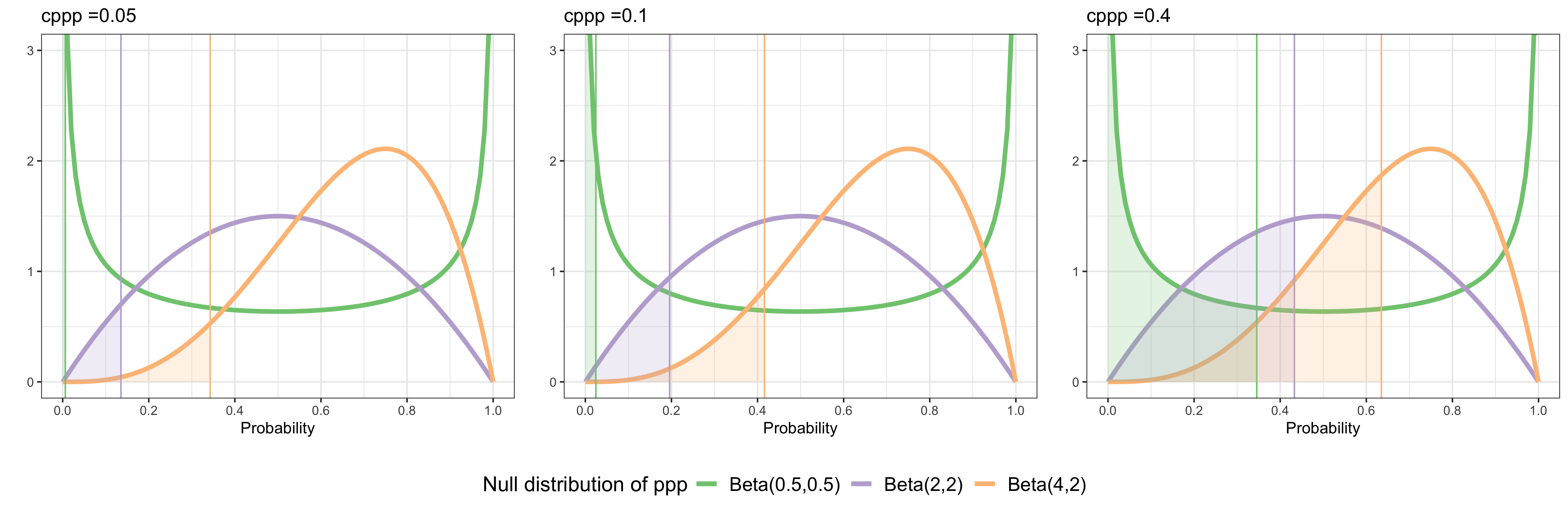}
  \caption{Examples of different scenarios for the null distribution of $\ppp(\Yc)$, for different choices of $\cppp(y)$. Each color shows a different null distribution. For a given null distribution, the vertical lines show the $\ppp$ value (on the x-axis) corresponding to the given $\cppp$ value for each sub-figure. For example, if (uncalibrated) $\ppp$ values really follow a Beta(2,2) distribution, and the $\cppp$ is really 0.1, then the $\ppp$ value will be about 0.2. The $\cppp$ procedure in effect uses Monte Carlo draws (via posterior simulation and MCMC runs) from an overdispersed $\ppp$ distribution (Beta-Binomial in these cases) to estimate the area under the curve to the left of 0.2.}\label{figSM0:example_cppp_scenario}
\end{figure}

\subsection*{Supplement to Section~4}

\begin{figure}
  \begin{center}
  \includegraphics[width=\textwidth]{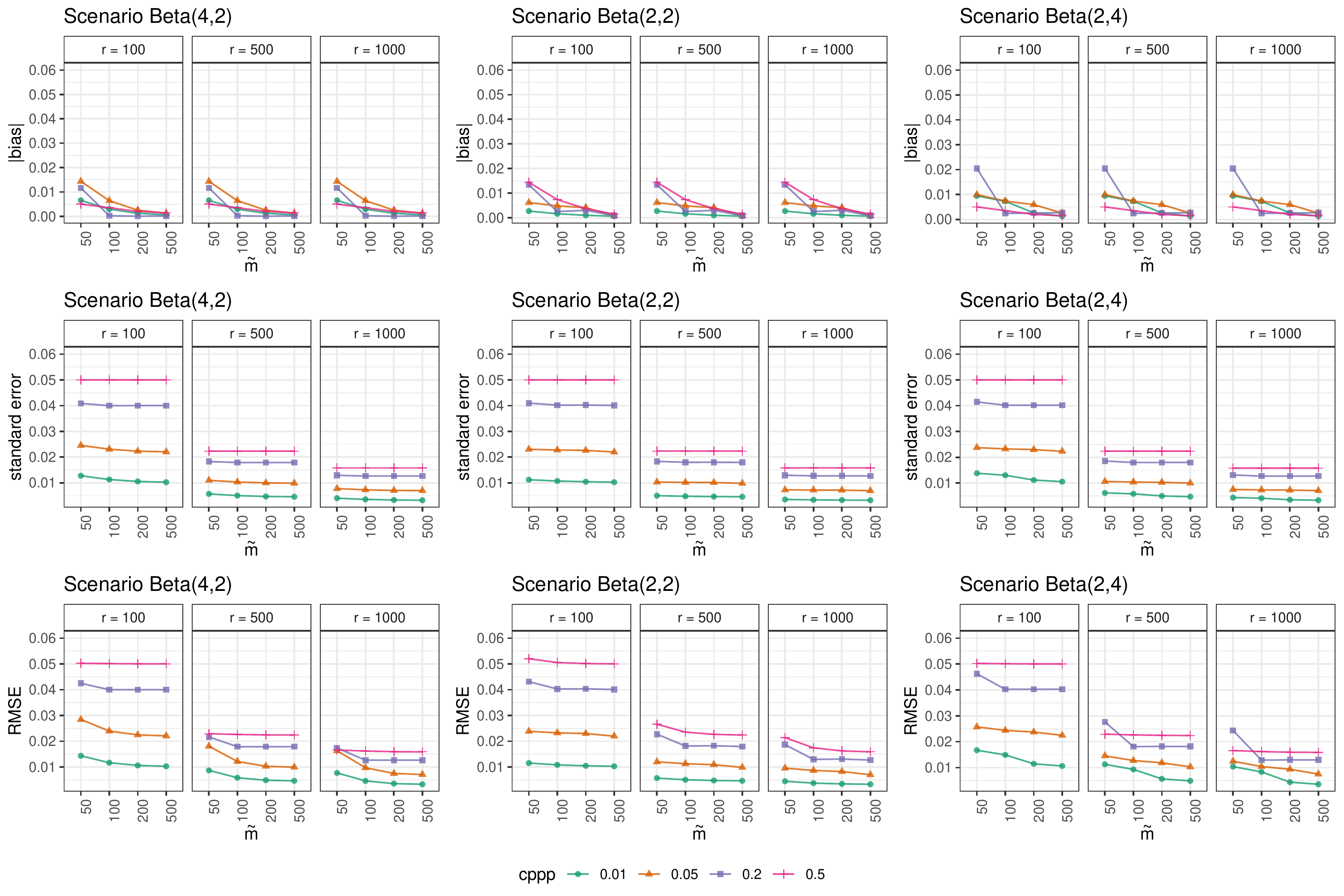}
  \end{center}
  \caption{Absolute bias (top row), standard error (mid row) and RMSE (bottom row) for $\cppph(y)$ considering different values of $\cppp(y) \in \{0.01, 0.05, 0.20, 0.50\}$ and combinations of $\mc$ and $r$. We consider two different scenarios for the null distribution of $\ppp(\Yc)$: in the first column $\ppp(\Yc) \sim \mbox{Beta}(4, 2)$ while in the second $\ppp(\Yc) \sim \mbox{Beta}(2,2)$.}
  \label{figSM1:beta_scenario_fixedM} 
\end{figure}

\begin{figure}
  \begin{center}
  \includegraphics[width=0.45\textwidth]{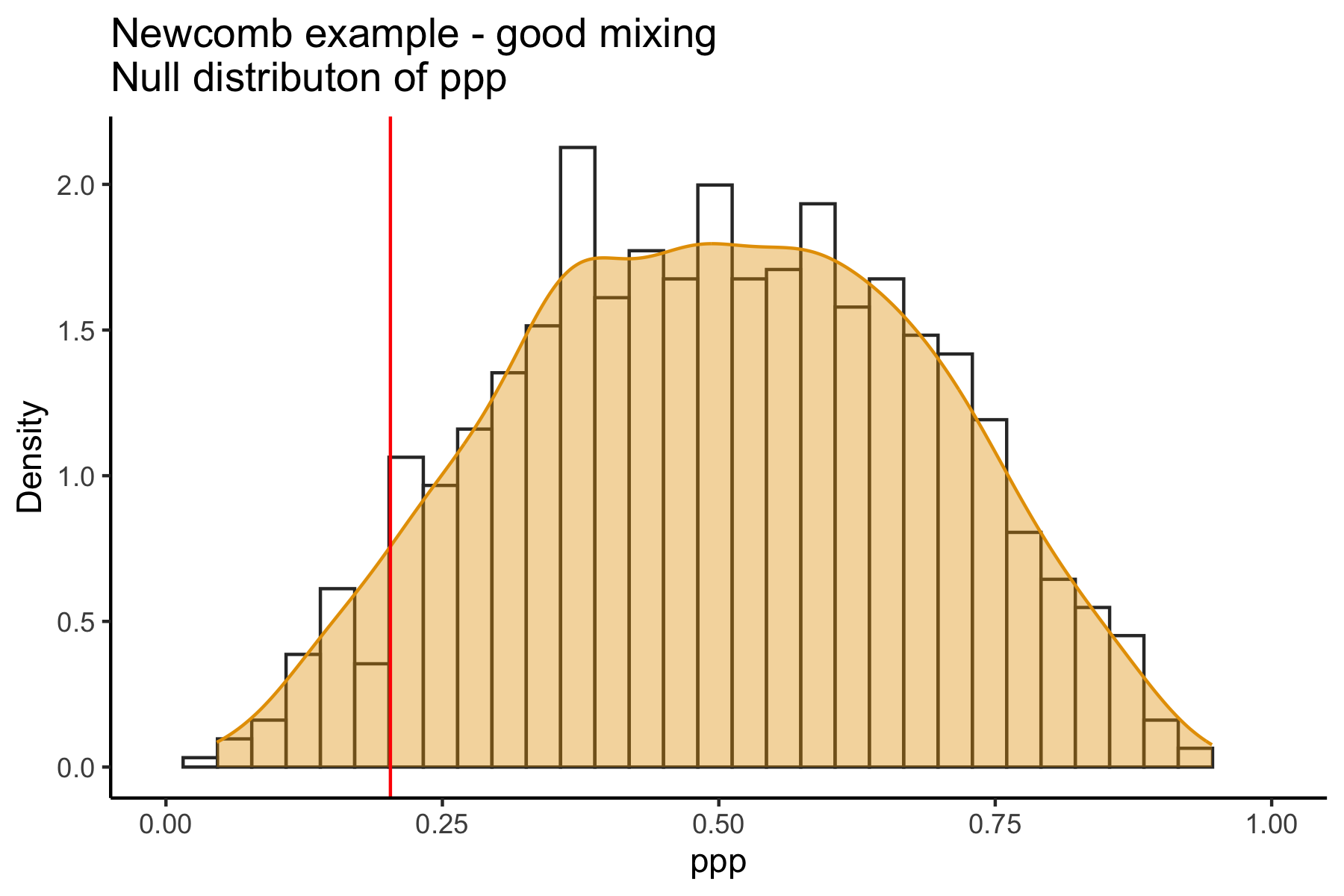}
  \includegraphics[width=0.45\textwidth]{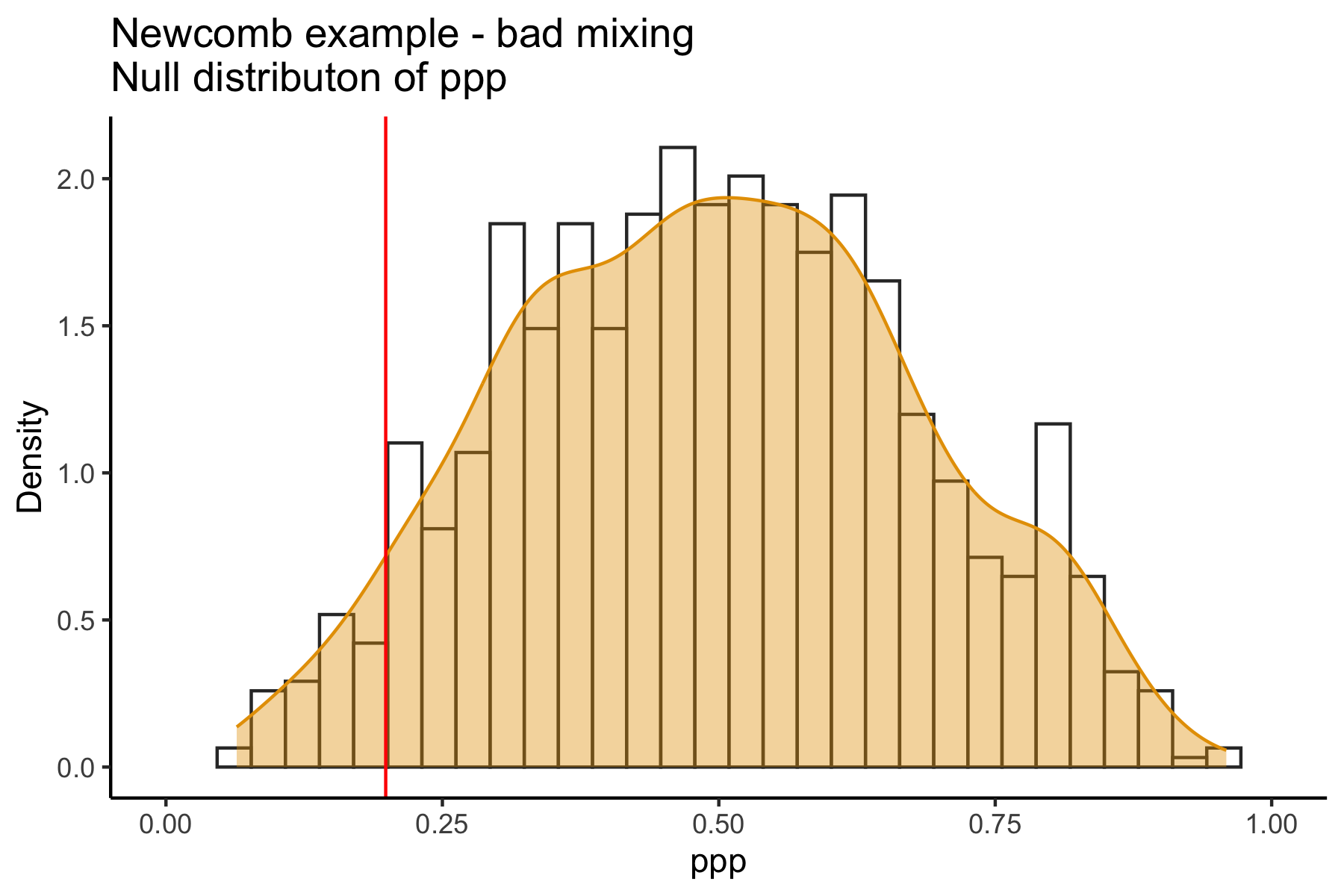}
\includegraphics[width=0.45\textwidth]{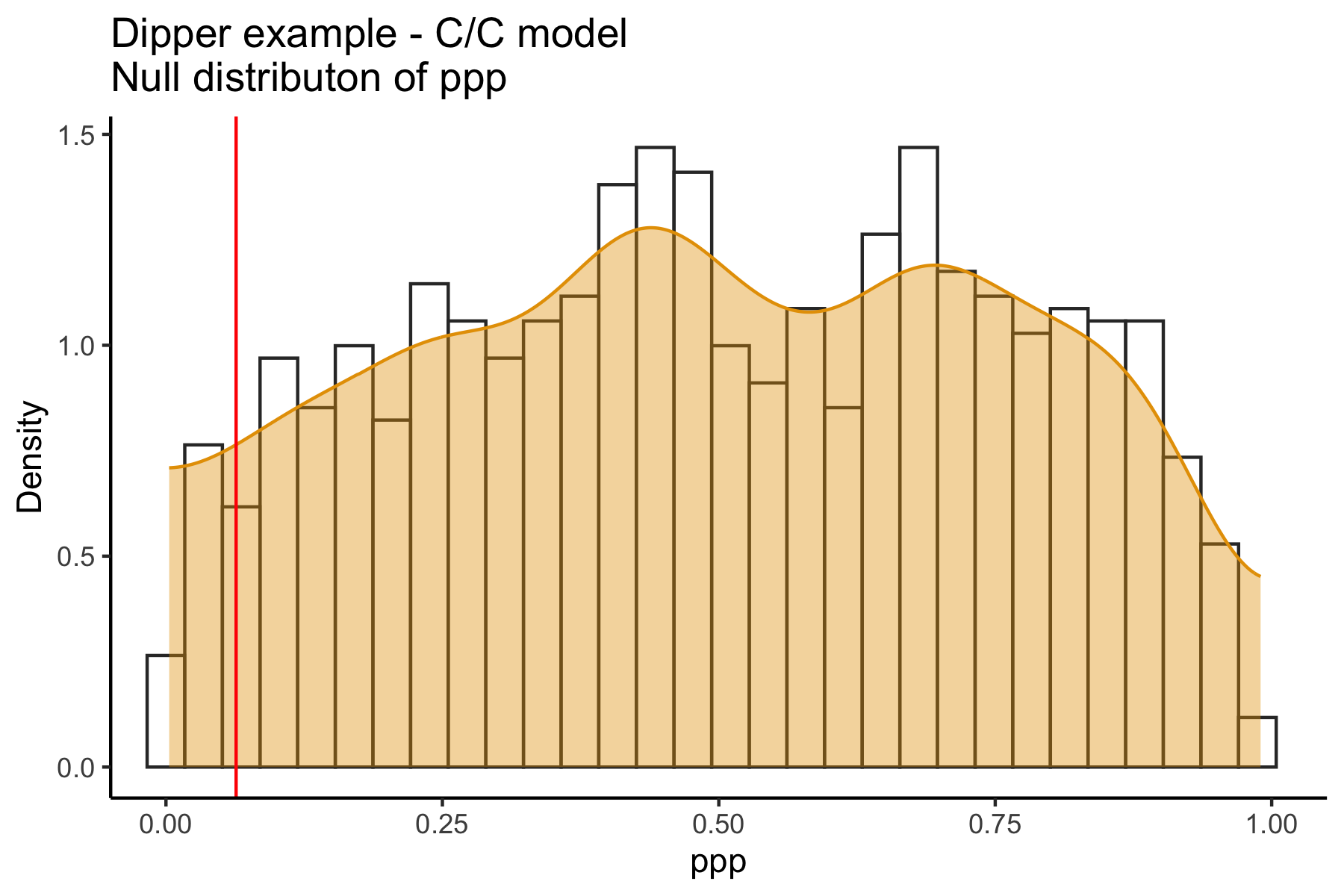}
  \includegraphics[width=0.45\textwidth]{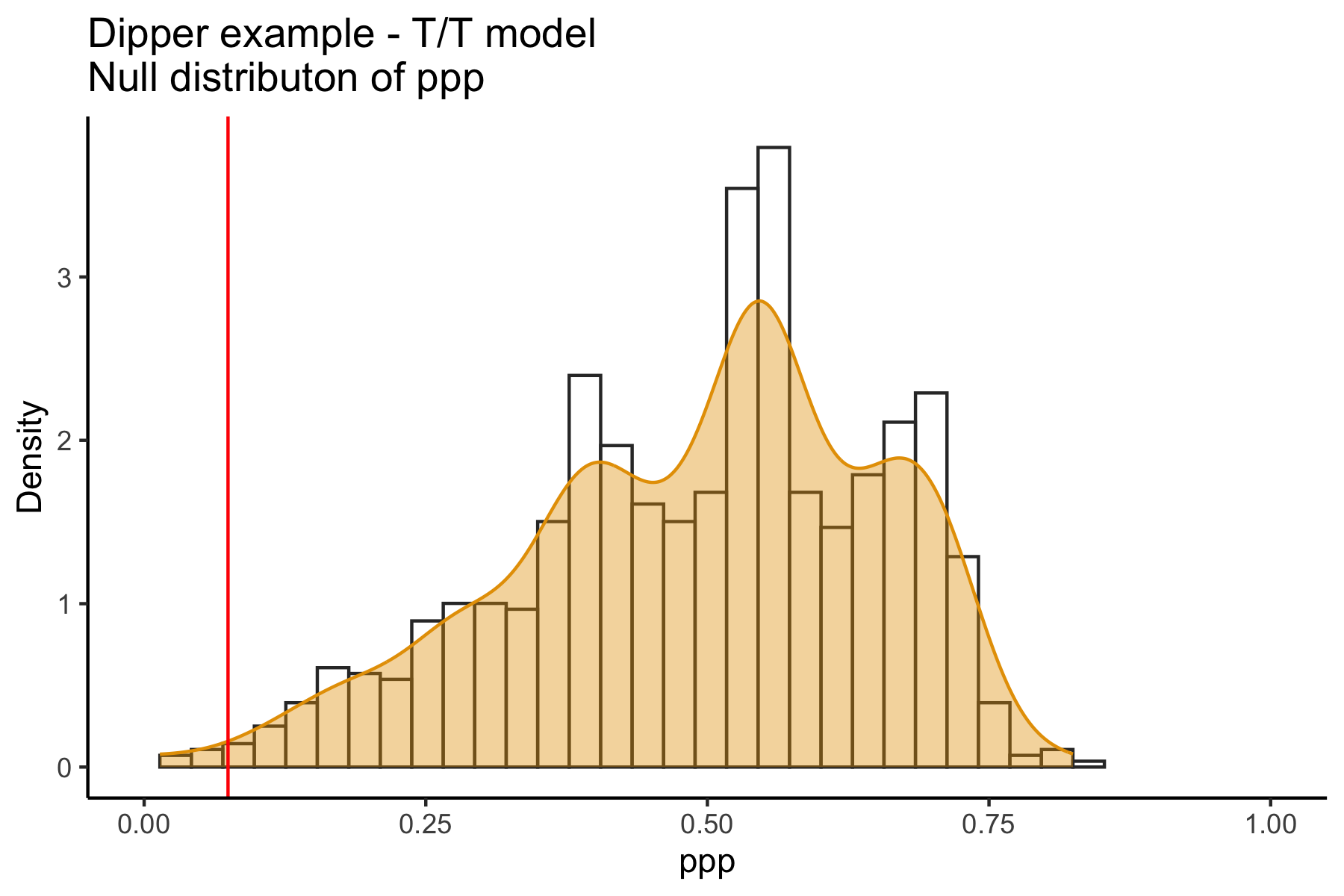}
  \includegraphics[width=0.45\textwidth]{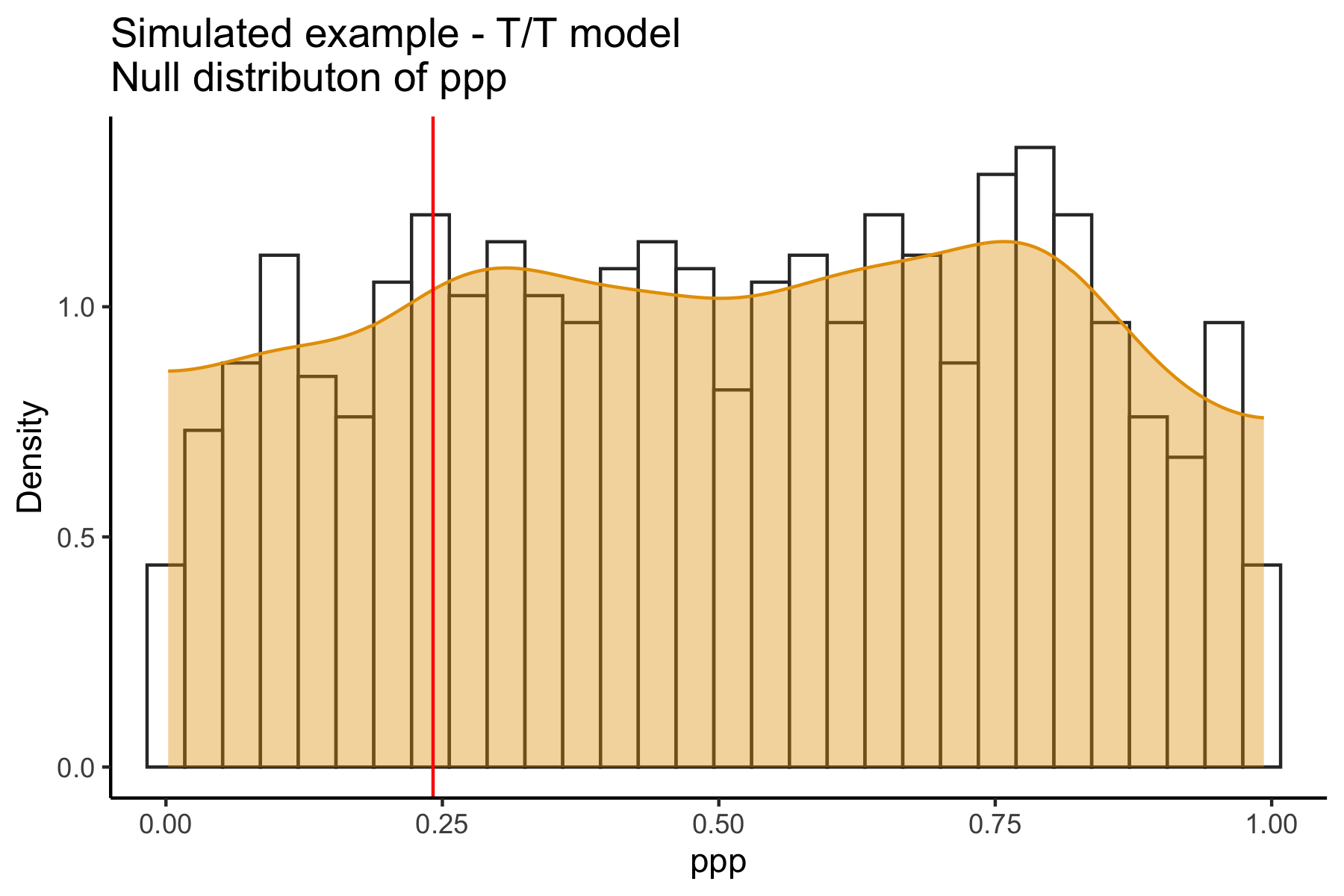}

  \end{center}
  \caption{Histogram and density plot of estimated ppps obtained using $r = m = 1,000$ for all the examples in Section~6. The red line indicates the observed value $\ppp(y)$.}
  \label{fig:ppp_distr}
\end{figure}

\inlineRevised{In Figure~\ref{figSM1:beta_scenario_fixedM} we show absolute bias, standard error, and root-mean-squared error for fixed values of $\mc \in \{50,100,200\}$ and number of calibration replicates $r \in \{100, 500, 1000, 2000\}$. Note that some of the results are equivalent to those in Figure~4 of the manuscript. }

\newpage
\subsection*{Supplement to Section~6}

\inlineRevised{For each of the examples in Section 6, Table~\ref{tab:times} report the times (in seconds) for the ppp and cppp computation. For the ppp computation we report the time taken for the original analysis and number of MCMC iteration used, while for the cppp we report times for different computational costs (combinations of $r$ and $\mc$). Note that we did not implement parallel computation for the calibration. Finally, we report the estimated time that a naive cppp would take when considering $r = 1,000$ calibration replicates and $\mc$ equal to number of MCMC iterations used in the original analysis. Computation of the ppp and cppp has been performed using a Linux cluster with 4 nodes having 24 cores and 128 GB RAM per node (Intel(R) Xeon(R) CPU E5-2643 v2 @ 3.50GHz).}

\inlineRevised{As the time required for MCMC sampling increases linearly with the number of iterations, t=the time for the cppp procedure ($t_{\cppp}$) can be approximated as a fraction of the original analysis time ($t_{\ppp}$), i.e. $t_{\cppp}(r, \mc) = (r \mc/m)t_{\ppp}$, without considering parallel computation for the calibration. This allows a straightforward approximation of required run-times for different computational budgets.}

\begin{table}[h]
\resizebox{\textwidth}{!}{%
\centering
\begin{tabular}{|l|cc|cccc|c|}
\hline
  \multicolumn{1}{|l|}{\bf{Example name}}&
  \multicolumn{2}{|c|}{\bf{original analysis (ppp)}}&
  \multicolumn{4}{|c|}{\bf{calibration (cppp)}} &
  \multicolumn{1}{|l|}{\bf{naive cppp}} \\ 
\hline
   & MCMC samples & time (sec) & 
  $c = 5,000$ & $c = 10,000$ & $c = 20,000$ & $c = 50,000$ & $r = 1,000$\\
  \hline
  Newcomb example - good mixing  & 5000  &  0.50 &  0.50 &  0.99 &   1.99 &   4.97 &    497 \\
  Newcomb example - bad mixing   & 5000  &  0.50 &  0.50 &  0.99 &   2.00 &   4.99 &    499 \\
  Dipper example -  C/C model    & 10000 & 83.05 & 41.52 & 83.05 & 166.10 & 415.25 &  83050 \\
  Dipper example -  T/T model    & 10000 & 50.42 & 25.21 & 50.42 & 100.85 & 252.12 &  50424 \\
  Simulated example -  T/T model & 10000 & 41.51 & 20.75 & 41.51 &  83.02 & 207.56 &  41512 \\
  \hline
\end{tabular}
}
\caption{Times in seconds for the original analysis (which computes the ppp) and calibration procedure (to compute the cppp) at different computational costs. For the original analysis, we also report the number of MCMC iterations used. In the last column, we report an estimate of the time that a naive implementation of the cppp would take. }
\label{tab:times}
\end{table}

\begin{table}
\resizebox{\textwidth}{!}{%
\centering
\begin{tabular}{|l|cccc|cccc|cccc|cccc|}
  \hline
  \multicolumn{1}{|l|}{\bf{computational cost}}& 
  \multicolumn{4}{|c|}{\bf 5,000} &
  \multicolumn{4}{|c|}{\bf 10,000} &
  \multicolumn{4}{|c|}{\bf 20,000} &
  \multicolumn{4}{|c|}{\bf 50,000} \\
  \hline
  \textbf{Calibration replicates} $r$   & 100 & 50 & 25 & 10
        & 200 & 100 & 50 & 20 
        & 400 & 200 & 100 & 40 
        & 1000 & 500 & 250 & 100 \\
  \textbf{MCMC samples} $\mc$ & 50 & 100 & 200 & 500 
        & 50 & 100 & 200 & 500
        & 50 & 100 & 200 & 500
        & 50 & 100 & 200 & 500 \\
  \hline
  \multicolumn{17}{|c|}{\bf Newcomb example - good mixing} \\
  \hline
  \textbf{Average ESS} & 49 & 99 & 198 & 496 & 49 & 99 & 198 & 496 & 49 & 99 & 198 & 496 & 49 & 99 & 198 & 496 \\
  \hline
  Plug-in & 0.988 & 0.970 & 0.774 & 0.568 
          & 0.998 & 0.958 & 0.952 & 0.708  
          & 1.000 & 0.982 & 0.930 & 0.886 
          & 1.000 & 0.984 & 0.982 & 0.938 \\
  Bootstrap - MBB & 0.988 & 0.982 & 0.890 & 0.556 
                  & 0.970 & 0.982 & 0.976 & 0.784 
                  & 0.940 & 0.978 & 0.970 & 0.924 
                  & 0.780 & 0.926 & 0.978 & 0.966 \\
  Bootstrap - Normal & 0.942 & 0.960 & 0.844 & 0.594 
                     & 0.874 & 0.974 & 0.960 & 0.740 
                     & 0.628 & 0.984 & 0.966 & 0.892 
                     & 0.000 & 0.962 & 0.968 & 0.946 \\
  \hline
  \multicolumn{17}{|c|}{\bf Newcomb example - bad mixing} \\
  \hline
  \textbf{Average ESS} & 20 & 41 & 81 & 204 & 20 & 41 & 82 & 203 & 20 & 41 & 82 & 203 & 20 & 41 & 82 & 203\\
  \hline
  Plug-in & 0.984 & 0.966 & 0.850 & 0.636 
          & 0.964 & 0.984 & 0.930 & 0.752 
          & 0.912 & 0.992 & 0.956 & 0.888 
          & 0.000 & 0.996 & 0.986 & 0.954 \\
  Bootstrap - MBB & 0.918 & 0.980 & 0.858 & 0.566 
                  & 0.794 & 0.962 & 0.944 & 0.788 
                  & 0.506 & 0.950 & 0.978 & 0.926 
                  & 0.000 & 0.868 & 0.986 & 0.960 \\
  Bootstrap - Normal & 0.546 & 0.950 & 0.886 & 0.670 
                     & 0.174 & 0.878 & 0.960 & 0.858 
                     & 0.002 & 0.686 & 0.974 & 0.918 
                     & 0.000 & 0.162 & 0.950 & 0.958 \\  
  \hline
  \multicolumn{17}{|c|}{\bf Dipper example -  C/C model} \\
  \hline
  \textbf{Average ESS} & 19 & 39 & 77 & 197 & 19 & 39 & 77 & 196 & 19 & 39 & 78 & 196 & 19 & 39 & 77 & 196\\
  \hline
  Plug-in & 0.996 & 0.970 & 0.884 & 0.610 
          & 0.998 & 0.996 & 0.928 & 0.724 
          & 1.000 & 1.000 & 0.964 & 0.836 
          & 1.000 & 1.000 & 0.984 & 0.940 \\
  Bootstrap - MBB & 0.962 & 0.976 & 0.842 & 0.462 
                  & 0.924 & 0.958 & 0.970 & 0.700 
                  & 0.802 & 0.968 & 0.986 & 0.904
                   & 0.248 & 0.974 & 0.984 & 0.950 \\
  Bootstrap - Normal & 0.720 & 0.980 & 0.908 & 0.636 
                     & 0.256 & 0.898 & 0.962 & 0.866 
                     & 0.014 & 0.724 & 0.976 & 0.910 
                     & 0.000 & 0.262 & 0.950 & 0.968 \\
  \hline
  \multicolumn{17}{|c|}{\bf Dipper example -  T/T model} \\
  \hline
  \textbf{Average ESS} & 13 & 25 & 49 & 120 & 13 & 25 & 49 & 119 & 13 & 25 & 49 & 120 & 13 & 25 & 49 & 120 \\
  \hline
  Plug-in & 0.994 & 1.000 & 0.958 & 0.352 
          & 0.970 & 1.000 & 0.970 & 0.552 
          & 0.864 & 0.990 & 0.982 & 0.636 
          & 0.000 & 0.962 & 0.972 & 0.776 \\
  Bootstrap - MBB & 0.882 & 0.938 & 0.672 & 0.164   
                  & 0.534 & 0.970 & 0.880 & 0.350 
                  & 0.054 & 0.868 & 0.984 & 0.622 
                  & 0.000 & 0.350 & 0.846 & 0.860 \\
  Bootstrap - Normal & 0.256 & 0.926 & 0.902 & 0.214 
                     & 0.000 & 0.842 & 0.972 & 0.472 
                     & 0.000 & 0.362 & 0.986 & 0.762 
                     & 0.000 & 0.000 & 0.722 & 0.834 \\
  \hline
  \multicolumn{17}{|c|}{\bf Simulated example - T/T model} \\
  \hline
  \textbf{Average ESS} & 44 & 88 & 175 & 441 & 44 & 88 & 176 & 444 & 44 & 88 & 175 & 439 & 44 & 88 & 175 & 440 \\
  \hline
  Plug-in & 0.964 & 0.962 & 0.918 & 0.912 
          & 0.986 & 0.968 & 0.932 & 0.932 
          & 0.998 & 0.984 & 0.958 & 0.946 
          & 1.000 & 0.998 & 0.980 & 0.962 \\
  Bootstrap - MBB & 0.968 & 0.970 & 0.936 & 0.900 
                  & 0.974 & 0.964 & 0.948 & 0.928 
                  & 0.986 & 0.968 & 0.962 & 0.934 
                  & 1.000 & 0.992 & 0.972 & 0.950 \\
  Bootstrap - Normal & 0.952 & 0.958 & 0.916 & 0.874 
                     & 0.964 & 0.964 & 0.944 & 0.916 
                     & 0.968 & 0.954 & 0.960 & 0.922 
                     & 1.000 & 0.982 & 0.962 & 0.934 \\
  \hline
\end{tabular}
}
\caption{
Confidence interval coverage for the examples in Section~\ref{sec:sec6} for each scenario of computational cost, number of calibration replicates and MCMC samples. We use normality approximation to calculate the confidence intervals at 95\% as $\cppph(y) \pm 1.96 \sqrt{ \widehat{\V}[\cppph(y)]}$ }.
\label{tab:coverage}
\end{table}

\newpage

\bibliography{bibliography}

\end{document}